\documentclass[preprint]{aastex}

\newcommand\na{$\nu^{-\alpha}$}
\newcommand\ssix{S$_{6}$}
\newcommand\dssix{$\delta$S$_{6}$}
\newcommand\stwenty{S$_{20}$}
\newcommand\dstwenty{$\delta$S$_{20}$}
\newcommand\da{$\delta$$\alpha$}

\slugcomment{To Appear in {\it Astrophysical Journal},
Volume 565, 1 February 2002}
\shorttitle{Pannuti et al.}
\shorttitle{Supernova Remnants in NGC 7793}

\begin{document}

\title{An X-ray, Optical and Radio Search for Supernova Remnants
in the Nearby Sculptor Group Sd Galaxy NGC 7793}

\author{Thomas G. Pannuti\altaffilmark{1}, Nebojsa Duric\altaffilmark{2},
Christina K. Lacey\altaffilmark{3},  
Annette M.N. Ferguson\altaffilmark{4}, Marcus A. Magnor\altaffilmark{5}
and Caylin Mendelowitz\altaffilmark{6}}

\altaffiltext{1}{MIT Center for Space Research, 77 Massachusetts Avenue, 
NE80-6015, Cambridge, MA 02139; tpannuti@space.mit.edu}
\altaffiltext{2}{Institute for Astrophysics, Department of Physics and 
Astronomy, University of New Mexico, 800 Yale Blvd., N.E., Albuquerque,
NM 87131}
\altaffiltext{3}{Department of Physics and Astronomy, University of
South Carolina, Columbia, SC 29208}
\altaffiltext{4}{Kapetyn Astronomical Institute, University of 
Groningen, Postbus 800, Groningen 9700 AV, The Netherlands} 
\altaffiltext{5}{Computer Graphics Laboratory, Department of Computer
Science, Stanford University, Gates Building, Room 360, Stanford,
CA 94305-9025} 
\altaffiltext{6}{Science Department, The Evergreen State College,
2700 Evergreen Parkway NW, Olympia, WA 98505} 

\begin{abstract}
This paper is the second in a series devoted to examining the
multi-wavelength properties of supernova remnants (SNRs) located
in nearby galaxies. We consider here the resident SNRs
in the nearby Sculptor Group Sd galaxy NGC 7793. Using our own 
Very Large Array (VLA) radio observations at 6 and 20~cm, as
well as archived $\it{ROSAT}$ X-ray data, previously published 
optical results and our own H$\alpha$ image, we have searched
for X-ray and radio counterparts to previously-known optically-identified
SNRs, and for new previously unidentified SNRs at these two 
wavelength regimes. 
Consistent with our prior results for NGC 300, only a tiny minority
of the optically-identified SNRs have been found at another wavelength.
The most noteworthy source in our study is N7793-S26, which is the only
SNR in this galaxy that is detected at all three wavelengths (X-ray,
optical and radio). It 
features a long ($\sim$ 450 parsecs) filamentary morphology that is 
clearly seen in both the optical and the radio images. N7793-S26's radio 
luminosity exceeds   
that of the Galactic SNR Cas A, and based on equipartition calculations we
determine that an energy of at least 10$^{52}$ ergs is required to
maintain this source. Such a result 
argues for the source being created by multiple supernova explosions
rather than by a single supernova event. A second optically-identified
SNR, N7793-S11, has detectable radio emission but is not detected in
the X-ray. 
A radio-selected sample of candidate SNRs has also been prepared by searching
for coincidences between non-thermal radio sources and regions of
H$\alpha$ emission in this galaxy, and this search has yielded five
new candidate radio SNRs, to be added to the 28 SNRs that have already
been detected by optical methods. A complementary search for new
candidate X-ray SNRs has also been conducted by searching for 
soft-spectrum sources ($\it{kT}$ $<$ 1 keV) that are coincident with
regions of H$\alpha$ emission. That search has yielded a candidate
X-ray SNR which is coincident with one (and possibly two) of the candidate 
radio SNRs, but considerable diffuse X-ray emission throughout the disk
of NGC 7793 reduces the efficacy of the search. 
Like NGC 300, very little overlap in identifications is seen between  
the SNRs found
through X-ray, optical and radio methods, and such a result argues
for the role played by distance-dependent selection effects in 
determining the detectability of SNRs. In addition, we find that
the density of the ambient interstellar medium (ISM) surrounding SNRs
significantly impacts the spectral characteristics of the
SNRs in this galaxy, consistent with surveys of the SNR populations
in other galaxies. 
\end{abstract}
\section{Introduction}
This paper is the second in a series devoted to the multi-wavelength study 
of supernova remnants (SNRs) in nearby galaxies. In our previous paper 
\citep*[hereafter referred to as Paper I]{Pannuti00} 
we analyzed observations made at the X-ray, optical and radio  
wavelengths of the nearby Sculptor Group Sd galaxy NGC 300. We sought to  
determine the X-ray and radio properties of the 28 SNRs identified 
previously in that galaxy through optical search techniques, namely 
H$\alpha$ and [\ion{S}{2}] narrow-band imaging \citep*[hereafter referred
to as BL97]{DDB80, BL97}, and in addition, we searched for new candidate 
SNRs at the 
X-ray and radio wavelengths to complement this prior optical work.  
Our search yielded sixteen new candidate X-ray and radio SNRs, later
reduced to fifteen by the recent work of \citet{RP01}, and the total
number of SNRs and candidate SNRs in NGC 300 is now 43. 
We found very little overlap between the three sets of selected candidates,
and we interpret this to indicate that a multiple-wavelength approach
is necessary to detect a maximum number of candidate SNRs in a 
particular galaxy. We also hypothesized that the limited overlap between the
selected sets of candidate SNRs indicated selection effects inherent in each
type of survey: optical surveys are biased toward
the detection of SNRs located in regions of low density and corresponding
low optical confusion, while X-ray and radio surveys have the 
opposite bias and favor the detection of SNRs located in regions of
high density. 
\par 
In this paper, we examine another Sculptor Group Sd galaxy, NGC 7793,   
and once again we consider observations made at the X-ray,  
optical and radio wavelengths. Following the paradigm set by our previous 
work (\citeauthor{Pannuti00}), our intent is to determine the X-ray and radio
properties of the previously-known
optically-identified SNRs as well as search for new candidate SNRs
at those two wavelengths.
Salient properties of NGC 7793 are listed in Table \ref{7793Props}:       
\citet{PC88} measured a distance to this galaxy  
of 3.38 Mpc and classified it as a member of the Sculptor Group of 
galaxies. Because of its proximity and its low inclination angle of 
50$^{\circ}$ \citep{T88}, this galaxy makes an 
excellent choice for the study of galactic properties. 
These studies include an optical survey for resident SNRs (\citeauthor{BL97}),  
analyses of its HI content \citep{CP90}, its surface photometry \citep{C85}
and its radio continuum properties \citep{H86}. Its X-ray properties
have also been the subject of X-ray analysis, based on observations performed
with the $\it{Einstein}$ satellite \citep{F92} and later observations
performed with the $\it{ROSAT}$ satellite \citep*[hereafter referred to
as RP99]{RP99}. In Section 
\ref{NGC7793ObsSection}, we describe the observations of this galaxy and
data reduction at each wavelength, beginning with the radio (Section 
\ref{NGC7793RadioObsSubSection}), followed by the optical (Section
\ref{NGC7793OptObsSubSection}) and concluding with the X-ray (Section 
\ref{NGC7793XrayObsSubSection}). We discuss the multi-wavelength properties of 
the optically-selected SNRs in Section \ref{NGC7793OptSNRsSection}, and   
then present the new candidate SNRs selected at the radio and X-ray  
wavelengths in Sections \ref{NGC7793RadioSNRsSection} and 
\ref{NGC7793XraySNRsSection}. A discussion of our findings in this work 
is presented in Section \ref{DiscussionSection} and finally our  
conclusions are given in Section \ref{ConclusionsSection}. 
\section{Observations and Data Reduction\label{NGC7793ObsSection}}
\subsection{Radio Observations and Data 
Reduction\label{NGC7793RadioObsSubSection}}
We observed NGC 7793 with the Very Large Array (VLA) of the National 
Radio Astronomy Observatory (NRAO\footnote{The National Radio Astronomy 
Observatory (NRAO) is a facility of the National Science Foundation 
operated under 
cooperative agreement by Associated Universities, Inc.}) at a wavelength
of 6~cm (4885 MHz) in the hybrid CnB configuration (northern arm in B 
array) on 1993 May 22 and at a wavelength of 20~cm (1465 MHz) in the
hybrid BnA configuration (northern arm in A array) on 1998 June 13. 
The greater north-south coverage provided by the extended northern arms
in these two arrays make them well-suited for studying such southern
galaxies as NGC 300 (\citeauthor{Pannuti00}) 
and NGC 7793 (the present paper). The approximate 
beam sizes in the final images are $\approx$ 6$\arcsec$ at 6~cm (4885 MHz) 
and $\approx$ 6$\arcsec$ at 20~cm (1465 MHz), and the RMS noise measured 
in each image is 32 $\mu$Jy/beam and 60 $\mu$Jy/beam at 6~cm and 20~cm,
respectively. Details of the VLA observations are provided in Table 
\ref{VLAObs}. 
\par
Our strategy for observing this galaxy was similar to the one described
in \citeauthor{Pannuti00} for observing NGC 300. Once again, to minimize
bandwidth smearing effects that may limit the field of view when observing 
in normal
continuum mode, we observed NGC 7793 in multichannel line mode with
eight channels per intermediate frequency (IF), including the continuum
calibrating channel (the zeroth channel), and a channel width of  
3.125 MHz. Of the seven spectral channels, the seventh was discarded because
its response decreased markedly, and the bandpass correction needed to
address this low response was both rather high and a possible source
of error. A total band of 37.5 MHz was synthesized after the seventh
channel was dropped. The effective observing frequencies for 
the images made from the remaining six channels were 1.448 GHz and 4.860
GHz at 20~cm and 6~cm, respectively. Data reduction and analysis were
performed using the AIPS software package, which is a standard data
reduction package provided by the NRAO. The final images were corrected
for primary beam attenuation before undertaking flux measurements.
\par
For the analysis of the radio sources in this galaxy, we repeated 
the procedure described previously in \citeauthor{Pannuti00}: again, our
primary goal was to find and measure all discrete
sources above a minimum 3 $\sigma$ detection level at both 6 and 20~cm.
Peak positions of the identified sources were determined by fitting
single Gaussians to their emission profiles at 20~cm. 
Routines in AIPS were used to identify the flux density associated
with each source: by using annuli that were approximately 1$\arcsec$
in width, we determined the flux density based on the cumulative flux
density minus the average flux density at an outer annulus chosen to
represent the background level.
\par
The spectral index $\alpha$ of each source (defined by the convention
$S$$_{\nu}$~$\propto$ \na) was calculated from the measured 6~cm and 20~cm
flux densities, denoted as $S$$_6$~and $S$$_{20}$, respectively. The 
uncertainties in each spectral index were determined by propagation
of errors. Those radio sources that were coincident with regions of  
H$\alpha$ emission in NGC 7793 and featured non-thermal spectral indices 
formed the set of radio
sources from which the sample of candidate radio SNRs was
assembled (as described in Section \ref{NGC7793RadioSNRsSection}).
For our selection criterion, we classified those radio sources with
spectral indices $\alpha$ $\geq$ 0.2 as candidate radio SNRs. This 
criterion is consistent with the previous searches for candidate radio
SNRs in the galaxies NGC 300 (\citeauthor{Pannuti00}) and NGC 6946
\citep{LDG97}.
Properties of these sources (positions in J2000.0 coordinates, flux
densities at 6 and 20~cm and spectral index $\alpha$) are listed in
Table \ref{N7793RadioSNRs}.  
\subsection{Optical Observations and Data 
Reduction\label{NGC7793OptObsSubSection}}
NGC 7793 was observed and imaged using an H$\alpha$ filter  
by the 1.5 meter telescope at the Cerro Tololo 
InterAmerican Observatory (CTIO)\footnote{The CTIO is operated by
AURA, Inc., under cooperative agreement with the National Science 
Foundation as part of the National Optical Astronomy Observatories.}
on the nights of 1993 December 14-18 and 1994 September 28-October 2. The 
observational data and its analysis are presented in \citet{FWGH96}, 
and the reader is referred to that work for more information.  
\par 
NGC 7793 was observed as part of a study of the diffuse ionized
gas (DIG) found in this galaxy as well as another Sculptor Group
galaxy, NGC 247. The DIG was found to account for between 30\% and
50\% of the diffuse H$\alpha$ emission in this galaxy, compared to 
between 44\% and 54\% for NGC 300 as found by \citet{HWG96}. \citet{GWTH98}
have found that the DIG accounts for between 40\% and 50\% of the
H$\alpha$ emission from three other galaxies -- namely M51, NGC 5195 and
M81 -- and it appears that generally speaking, the DIG accounts for
approximately this fraction of H$\alpha$ emission within spiral galaxies.  
The presence of considerable DIG in NGC 7793 (especially when the DIG 
accounts for such a high fraction of a galaxy's diffuse H$\alpha$
emission) suggests that extensive star formation is taking place in
this galaxy. This current epoch of widespread star formation will
certainly produce a population of SNRs that will be detectable by
our multi-wavelength survey. 
\par
The 28 SNRs found in NGC 7793 by \citeauthor{BL97} were detected
through narrow-band imaging of this galaxy using H$\alpha$ and 
[\ion{S}{2}] filters. A particular source was identified as an SNR candidate
if the ratio of its fluxes [\ion{S}{2}]/H$\alpha$ exceeded 0.4. We
will examine the X-ray and radio properties of these sources in Section
\ref{NGC7793OptSNRsSection}. For the present work, we will also make use of 
the catalog of HII regions in this galaxy that was prepared by \citet{H69}, 
based on observations made using the Palomar 48-inch telescope. In that
paper, NGC 7793 was one of twenty galaxies that had its HII region
population surveyed and cataloged for future reference purposes. As a 
result of that survey, 35 HII regions were identified in this galaxy,
and we will use this list to identify HII regions that may be associated
with SNRs. We note that the survey presented in that paper was biased toward
the detection of only the brightest HII regions in NGC 7793, and for
that reason numerous faint HII regions were not listed.  

\subsection{X-ray Observations and Data 
Reduction\label{NGC7793XrayObsSubSection}}

We used archived observations of NGC 7793 made with the wide-angle,
spectral-resolving Position Sensitive Proportional Counter ($\it{PSPC}$) 
instrument aboard the $\it{ROSAT}$ satellite \citep{T92}. NGC 7793 
was observed over two separate epochs with the $\it{PSPC}$ instrument: the 
first epoch was on 1992 December 7 for 13332 seconds and the second was 
between 1993 May 20 and 1993 May 27 for 10507 seconds. Data from these two 
observing epochs were combined to produce images corresponding to a total 
integration time of 23839 seconds. Following the example set by \citet{L96} in 
their study of the X-ray sources in M33 and similar to our previous analysis 
of X-ray emission from NGC 300 (\citeauthor{Pannuti00}), images were 
prepared at the 
three energy bands defined as the ``total" (0.1-2.4 keV), the ``hard"
(1.0-2.4 keV) and the ``soft" (0.1-1.0 keV) from combining data for the 
two separate observation epochs. 
\par
\begin{figure}
\figurenum{1}
\epsscale{0.95}
\plotone{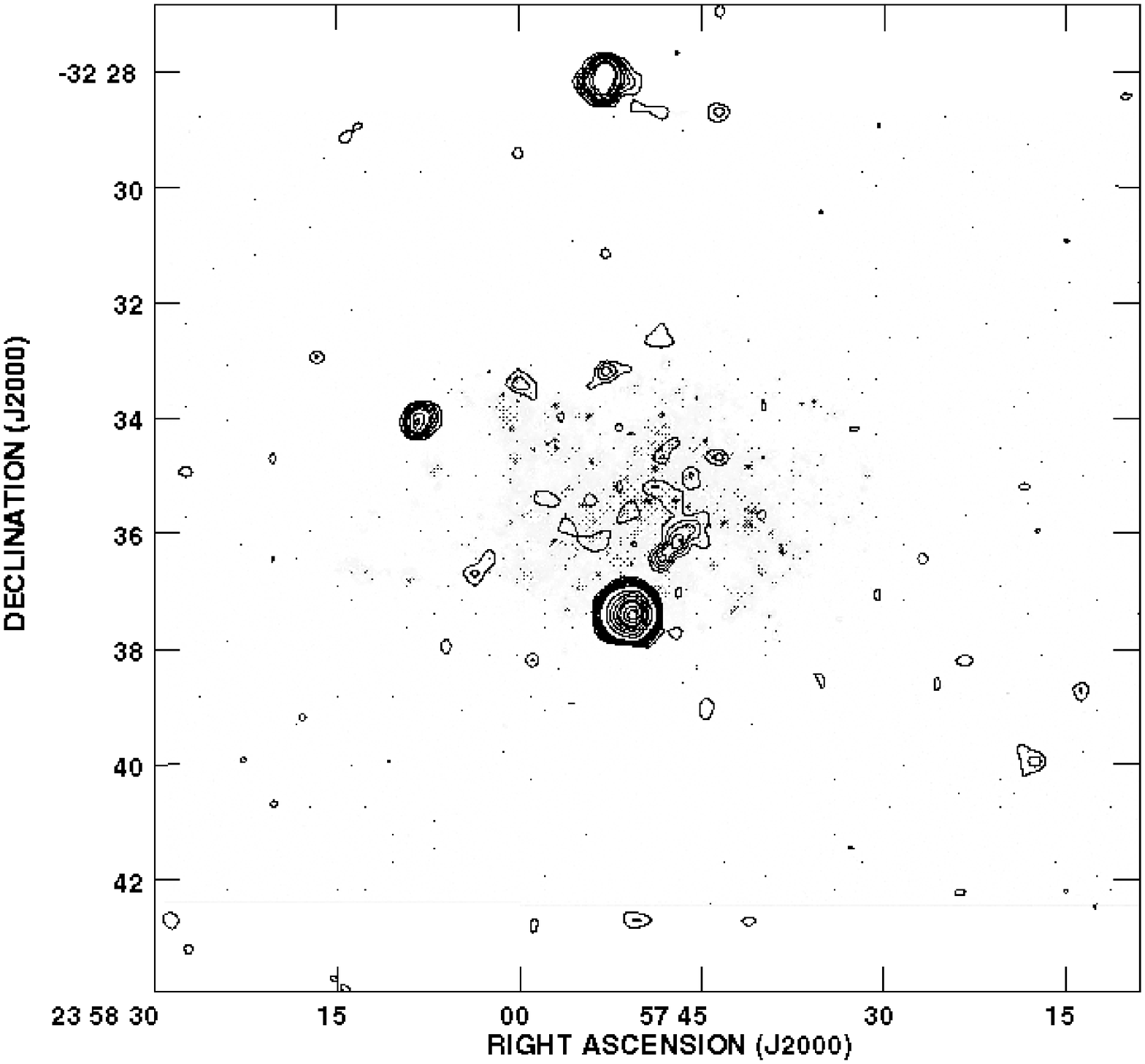}
\caption{X-ray (contours) and H$\alpha$ (gray-scale) image
of NGC 7793. The H$\alpha$ image was prepared using observations made
with the CTIO, while the X-ray image was prepared using observations
made with the ROSAT PSPC instrument over the total energy range of
0.1-2.4 keV with a combined integration time of 23839 sec. The noise level
in the X-ray image is 9.4 $\times$ 10$^{-4}$ counts per square arcminute,
and the contours are set at the 3$\sigma$, 4.5$\sigma$, 6$\sigma$, 
7.5$\sigma$, 9$\sigma$, 12$\sigma$, 14$\sigma$, 30$\sigma$, 45$\sigma$,
60$\sigma$, 75$\sigma$, 85$\sigma$, 105$\sigma$ and 120$\sigma$ 
level.\label{fig1}} 
\end{figure}

This $\it{PSPC}$ data of NGC 7793 has already been discussed in detail by 
\citeauthor{RP99}, and in their analysis 27 sources were identified at a 
4$\sigma$ detection level or higher in the central 25$\arcmin$ $\times$
25$\arcmin$ area of the field of view. Of these 27 sources, seven (denoted 
as P6, P7, P8, P9, P10,
P11 and P13 in their notation) fell within the optical extent of the galaxy.
We will concentrate on these particular seven X-ray sources to avoid
confusion with X-ray emission from background sources (such as distant
AGN's and quasars), and we present 
properties of these sources (source numbers, positions, net numbers of 
counts at the total band, fluxes, hardness ratios and probabilities of
variability), as measured by those authors,  
in Table \ref{NGC7793XraySources}. These fluxes given by 
\citeauthor{RP99} were determined by assuming a 5 keV thermal 
bremsstrahlung and a hydrogen column density of $N_H$ = 1.14 $\times$
10$^{20}$ cm$^{-2}$. The hardness ratios correspond to the particular
ratio denoted as ``HR1" by \citeauthor{RP99}, which is the ratio
defined as (hard counts - soft counts)/(hard counts + soft counts).   
In the last column, we list optical and  
radio counterparts to each X-ray source, drawing from the sample of
radio sources described in Section 2.1, the optically-identified SNRs
found by \citeauthor{BL97} and the catalog of HII regions found by
\citet{H69}. The HII regions are referred to by their catalog number; 
$\it{e.g.,}$ \#31 in that catalog is listed as H31. 
In Figure \ref{fig1}, the X-ray emission in the ``total" band 
from NGC 7793 (as observed by the $\it{ROSAT}$ $\it{PSPC}$) is depicted in 
contours, with H$\alpha$ emission shown in gray-scale. 
\par
The population
of X-ray sources within the disk of NGC 7793 is expected to be a mixture
of X-ray binaries and SNRs, and a spectral analysis can help differentiate 
between these two classes of sources. \citet{S99} has found that the      
spectra of X-ray binaries with low foreground column densities (which is
applicable for the present study of sources in NGC 7793) can be best fit
with power laws, while in contrast the X-ray spectra of SNRs is best fit
using thermal bremsstrahlung models \citep{HPH00}. Effects of thermal 
continuum absorption at radio wavelengths along Galactic lines of sight 
are discussed by \citet{LLKDBD01}: those authors discovered such absorption
toward the Galactic SNR W49B by an intervening complex of HII regions and
analyzed how this absorption affects the observed radio spectrum of the
SNR. Searching for variations in the luminosities of the X-ray sources  
detected in NGC 7793 gives an additional method for classifying these
sources as either X-ray binaries or SNRs:  
X-ray binaries can exhibit a time dependence
to their luminosities (as will other classes of variable X-ray sources), while
the X-ray emission from SNRs should be nominally time-independent.
By considering these differences in X-ray emission characteristics between
SNRs and X-ray binaries, we should be able to generally differentiate
between sources of these two types. We hasten to add, however, that
these distinctions cannot be viewed as absolute ways to classify these
sources. For example, \citet{S99} points out that a combination of
a power law model and a thermal bremsstrahlung model is required to 
optimize fits to the spectra of X-ray binaries, and likewise some 
SNRs are known to possess components to their X-ray emission which
have a power law dependence \citep{A97}. Finally, if the period
of the X-ray binary's emission exceeds the integration time for an 
X-ray observation of a galaxy, it may appear to be constant in X-ray
luminosity and lead to an incorrect classification. Nonetheless, we
will still make use of these general differences in the properties 
of X-ray emission from SNRs and X-ray binaries in order to approximately
classify these sources, and we note that more rigorous classifications
must await X-ray observations of this galaxy with superior quality. 
\par
We point out that at a distance of 3.38 Mpc \citep{PC88}, NGC 7793 is 
considerably more distant than NGC 300, which is located at a distance
of 2.1 Mpc \citep{FMHHMS92} and was the subject of our previous study 
(\citeauthor{Pannuti00}). For this reason, while our angular
resolution is the same in both of these studies, our spatial resolution
of structures in NGC 7793 is reduced. In addition, while only a modest
amount of diffuse X-ray emission was found in NGC 300 by \citet{RP01},
such emission was clearly detected in the 
disk of NGC 7793 by \citeauthor{RP99}, with
a characteristic energy $\it{kT}$ of approximately 1 keV. This diffuse
emission makes it difficult to clearly associate X-ray emission with 
point sources found at other wavelengths in this galaxy. We will return
to this point in the remaining sections of this paper. 
\section{Multi-Wavelength Properties of the Optically-Selected SNRs in 
NGC 7793\label{NGC7793OptSNRsSection}}
In our prior study of the optically-identified SNRs in NGC 300, we
found that most of the SNRs selected in this manner were weak or
undetected in the X-ray and radio (\citeauthor{Pannuti00}).
We hypothesized that such a result may indicate a selection effect
inherent in optical surveys, which are more sensitive to detecting
SNRs in regions of lower density in a galaxy. Such SNRs may be weak
emitters in the X-ray and radio because of their location within diffuse
environments, which are not conducive to high X-ray and radio 
luminosities. In contrast, SNRs which are powerful sources of X-ray
and radio emission are often located in regions of high density (such
as HII regions), and precisely because of their location they may
be missed by optical surveys, which are subject to considerable confusion
from diffuse emission in these regions. It is generally believed
that SNRs which are located in regions of low density were parented
by low-mass white dwarf progenitor stars as Type Ia supernovae (SNe), 
while in contrast SNRs which are found in regions of high density 
(such as spiral arms) were created by high-mass progenitor stars as 
Type II SNe \citep{B94}. However,  
exceptions have been pointed out for both of these assertions: as 
examples, we note that an isolated high-mass star may produce a 
Type II SNe in a region of comparatively low density, and a Type Ia
SNe may occur in a spiral arm of a galaxy (a location of enhanced
density) given the long lifetimes of white dwarf stars. 
\par
For our samples of extragalactic
SNRs, however, information about the type of SN that produced each SNR
is usually not available. Therefore, our studies of SNRs in other 
galaxies encompass SNRs produced by both low-mass and high-mass progenitor
stars and therefore sample much of the galaxy's star formation
history. We argue that by conducting surveys for SNRs in the X-ray,
optical and radio, we can probe regions of galaxies over a wide range
of densities and thus detect a maximum number of SNRs in galaxies, 
including SNRs that have been produced by both types of SNe and 
progenitor stars. 
\par
In our study of the optically-identified SNRs in NGC 7793, we discovered 
errors in the published positions of these SNRs as given by \citeauthor{BL97}. 
We found that while the published right ascension positions for the SNRs 
located in the eastern half of the galaxy appeared to be reasonably accurate 
-- that is, within the quoted positional errors of $\approx$ 2$\arcsec$
(\citeauthor{BL97}) -- the accuracy of the listed right ascension positions 
declined markedly for the SNRs
in the galaxy's western half. In particular, the accuracy in the published
right ascension of the SNRs was considerably beyond the 
quoted positional errors for the SNRs N7793-S1 through N7793-S12. 
For N7793-S12, we found a difference of approximately 19$\arcsec$,
and this difference increased gradually as we considered SNRs in turn
located toward the galaxy's western edge, culminating with 37$\arcsec$ for
N7793-S1. We note that the published declination positions were accurate
within the quoted positional errors for the entire sample of      
optically-identified SNRs. To determine the correct right ascension 
coordinates of the optically-identified SNRs, we measured the positions  
of the SNRs  
as seen in our H$\alpha$ image based on the images of these sources
as presented by \citeauthor{BL97}, and we give the corrected positions in 
Table \ref{N7793OptSNRsPositions}. After confirming that we had located the  
SNRs identified by \citeauthor{BL97},   
we proceeded to analyze their X-ray and radio properties.
\par
\begin{figure}
\figurenum{2}
\epsscale{0.95}
\plottwo{f2a.eps}{f2b.eps}
\plotone{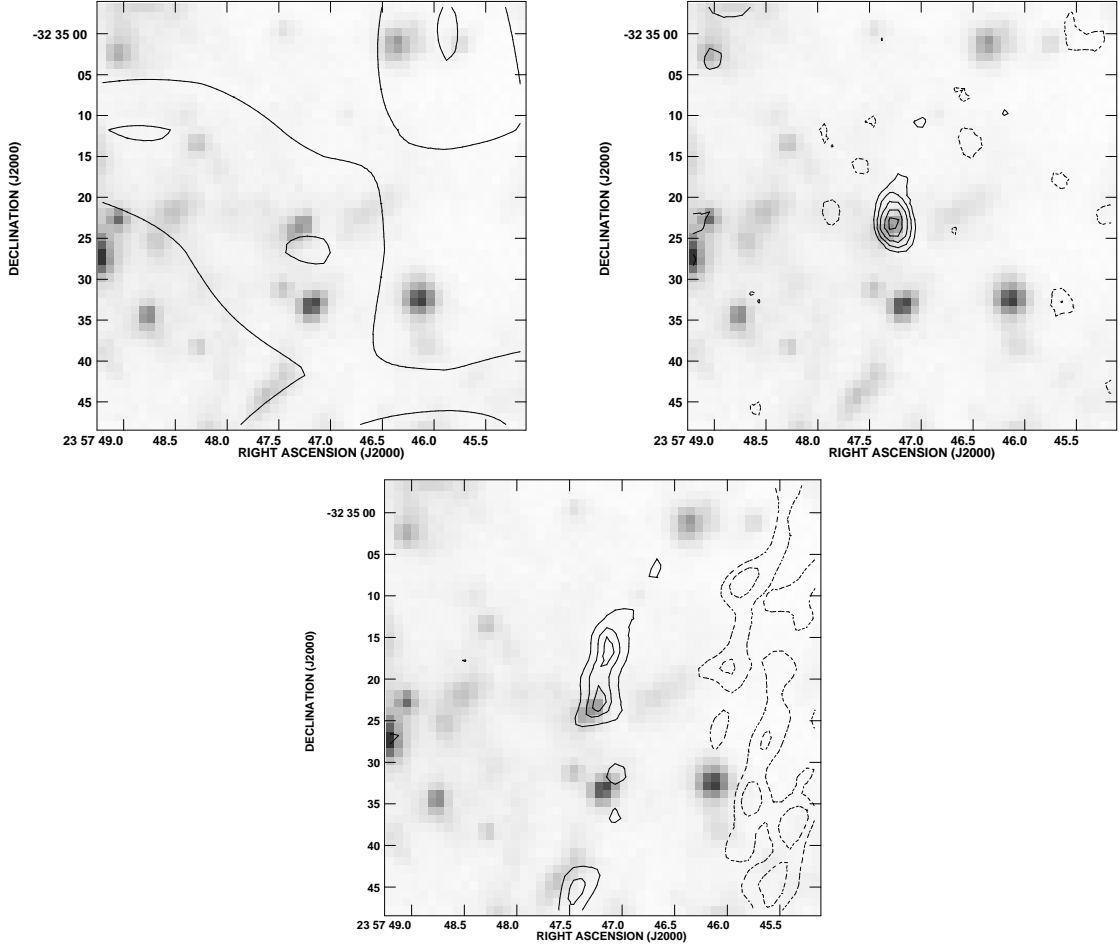}
\caption{Multi-wavelength images of the environment
surrounding the optically-identified SNR N7793-S11. In each image,
the H$\alpha$ emission is depicted in gray scale. X-ray emission
(as observed by the PSPC instrument) is depicted in Fig. 2a (the
upper left figure), using
the same contour levels as in Figure 1. Emission at the 6 and 20~cm
is depicted in contours in Fig. 2b and 2c (the upper right and lower
figures), respectively. In Fig. 2b,
the contours are placed at -3, -2, 3, 4, 5, 6 and 6.75 times the rms
noise level at 6~cm (32 $\mu$Jy), and in Fig. 2c, the contours are
placed at -3, -2, 3, 4 and 4.5 times the rms noise level at 20~cm
(60 $\mu$Jy). The approximate beam sizes are $\approx$ 6$\arcsec$ in
both the 6 and 20~cm images. See $\S$ 
\ref{NGC7793OptSNRsSection}. \label{fig2}} 
\end{figure}
\begin{figure}
\figurenum{3}
\epsscale{0.95}
\plottwo{f3a.eps}{f3b.eps}
\plotone{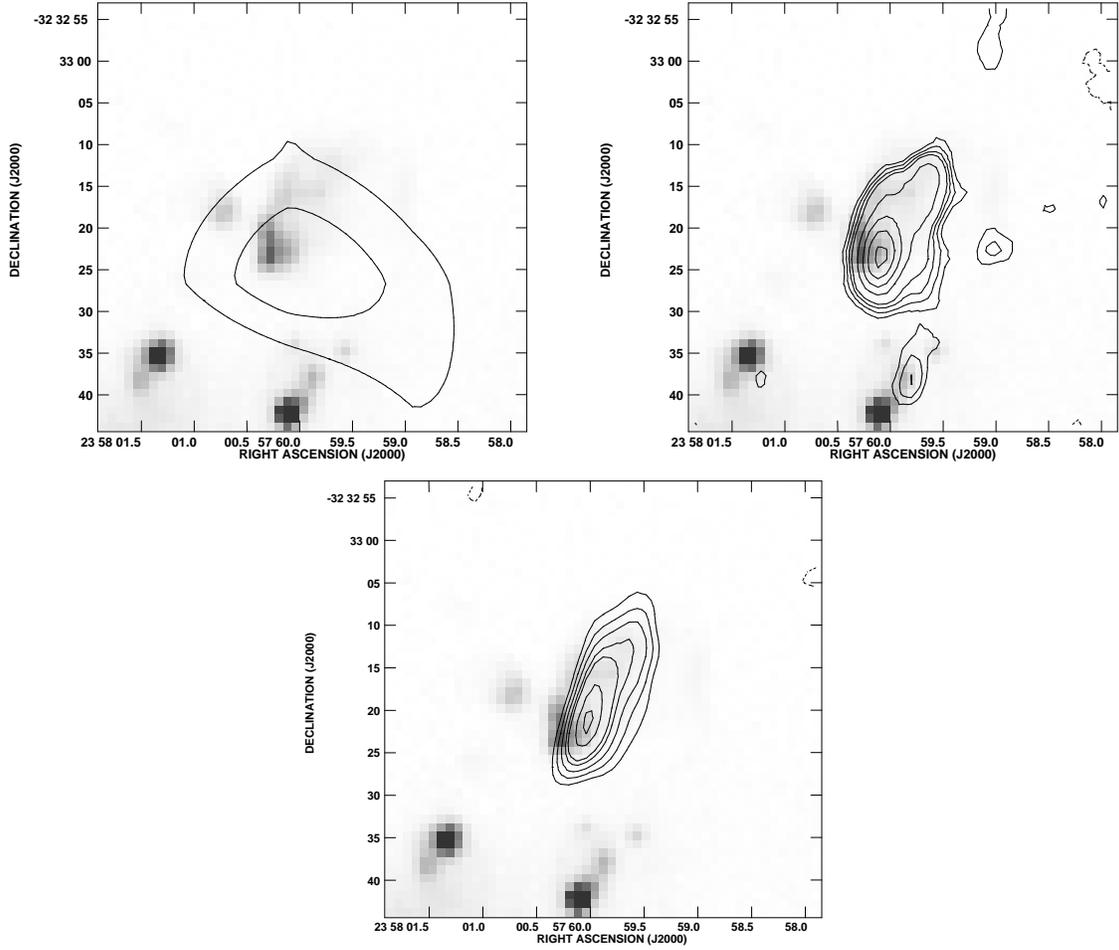}
\caption{Same as Fig. 2, but for the optically identified
SNR N7793-S26. Notice the filamentary structure apparent in both H$\alpha$
emission and radio emission, as well as the coincidence with the X-ray
source P8 as seen in Fig. 3a. 
In Fig. 3b, the contours are placed at -3, -2,
3, 4, 5, 6, 9, 12, 15 and 18 times the rms noise level at 6~cm, and in 
Fig. 3c, the contours are placed at -3, -2, 9, 12, 15, 18, 20, 25 and 28  
times the rms noise level at 20~cm. See $\S$ 
\ref{NGC7793OptSNRsSection}. \label{fig3}} 
\end{figure}

Of the 28 SNRs found in NGC 7793 by \citeauthor{BL97} using optical methods,
only two were detected at another wavelength at the 3 $\sigma$ level
or higher. N7793-S11 was detected as a non-thermal radio source with
flux densities of 0.27$\pm$0.06 mJy and 0.42$\pm$0.12 mJy at the 
wavelengths of 6 and 20~cm, respectively, corresponding to a
spectral index $\alpha$ of 0.4$\pm$0.2. No X-ray counterpart was
detected for this source, though a search for a counterpart is complicated
by the diffuse X-ray emission within the disk of  
this galaxy. We did find both X-ray and
radio counterparts to the source N7793-S26, an SNR which also
has a filamentary extension denoted as N7793-S26ext by \citeauthor{BL97}.
For the remainder of this paper, we will refer to both the source N7793-S26 
and its filamentary extension N7793-S26ext as N7793-S26. We present a 
listing of the spectral properties of these two optically-identified SNRs 
in Table \ref{N7793RadioSNRs}, and multi-wavelength images of these
two optically-identified SNRs are presented in Figures \ref{fig2} and 
\ref{fig3} respectively.  
\par
N7793-S26 is a particularly interesting source, and we will devote
some discussion to its properties here. It is readily apparent in
optical images that the source's morphology is that of a long filament,
and this morphology is imitated remarkably in the radio, as seen
by inspection of the images presented in Figure \ref{fig3}. 
\citeauthor{RP99} first identified the X-ray counterpart to N7793-S26 
source (labelled by those authors as P8) and attempted to fit its X-ray 
spectrum using two models (namely a thermal bremsstrahlung model and a
blackbody model). While the goodness of the model fits ($\chi^2$'s of 
0.21 and 0.16, respectively) do not strongly favor one model over 
the other, we do note that the fit parameter for the blackbody model 
returned an energy value of $\it{kT}$ = 0.17$\pm$0.05 keV, while the 
corresponding energy value for the thermal bremsstrahlung model is 
$\it{kT}$ = 0.36$\pm$0.25 keV.
Both of these energy values are consistent with that expected for the X-ray
emission from an SNR. In addition, \citeauthor{RP99} also performed a test
to search for variability in the emission from the X-ray sources in NGC 
7793, and found that the probability of variability in P8's X-ray 
emission was only 13\%, also consistent with the type of X-ray emission
expected from SNRs. 
We will discuss the variability of X-ray emission from
the X-ray sources in NGC 7793 more extensively in Section 
\ref{NGC7793XraySNRsSection}. 
\par
As part of their analysis, \citeauthor{RP99} examined an NRAO
VLA Sky Survey (NVSS) 1.4 GHz radio continuum image of NGC 7793, and
found a radio counterpart to N7793-S26 to complement the X-ray and 
optical detections.
This three-wavelength detection enhances the likelihood that this  
source is either a collection of several SNRs or a
single large and luminous SNR. No quantification of the radio emission
properties (such as flux densities) was given by \citeauthor{RP99} for this
source, but by examining our 6 and 20~cm images we measure flux densities
for the entire structure of 1.24$\pm$0.19 mJy and 3.75$\pm$0.30 mJy at 6 
and 20~cm, respectively, with a  
corresponding spectral index of $\alpha$=0.9$\pm$0.2. At the 
assumed distance of 3.38 Mpc for NGC 7793, the most radio-luminous
Galactic SNR, Cas A, would have a flux density of 2.8 mJy at 20~cm,
which is about 75\% of the observed flux density for 
N7793-S26 at that same wavelength. Clearly, this is an  
extremely luminous radio source, and its filament-like morphology with
a length of approximately 450 parsecs makes this source even more
intriguing.   
\par
The unusual properties of this source have led to speculation 
about its true nature. It is possible that this source was
produced by the collective expansion of several SNRs that were created 
at approximately the same time, therefore indicating that N7793-S26
may be a nascent superbubble structure. 
\citet{M97} argue that the maximum attainable size of an SNR expanding
into an interstellar medium (ISM) with a mean number particle density of
between 0.1 and 1 cm$^{-3}$ (a typical Galactic value) is approximately
100 parsecs. In addition, based on equipartition calculations 
where we assume a radius of 225 pc for N7793-S26, consistent with
its apparent size, we determine that an energy of at least 10$^{52}$
ergs is required to maintain this source (Pannuti et al. 2002,
in preparation), far greater than the energy yield expected
from a single SN. All of this evidence suggests that N7793-S26 was
not produced by a single supernova explosion of average energy. 
\par
\citeauthor{BL97} investigated the nature of N7793-S26 by searching
for an interior star cluster within the SNR. 
The presence of such a cluster would argue 
in favor of the collective supernova remnant hypothesis. Those authors
found no evidence for such an interior star cluster, but their data
lacked the precision required to answer this question decisively.
Therefore, the true nature of this object has yet to be determined, and it    
will likely be the subject of further study at multiple wavelengths.
\par
To summarize the
results of this section, of the 28 optically-identified SNRs detected
in this galaxy by \citeauthor{BL97}, only two possess counterparts at 
another wavelength: N7793-S11 in the radio and N7793-S26 in both the
X-ray and the radio. Consistent with our previous findings for
the SNR population in NGC 300 (\citeauthor{Pannuti00}), no X-ray or radio
emission has been found from the large majority of the optically-identified
SNRs in NGC 7793. The results of our complementary searches for candidate
radio and X-ray SNRs are presented in Sections \ref{NGC7793RadioSNRsSection}
and \ref{NGC7793XraySNRsSection}, respectively.  
\section{Properties of the Radio-Selected SNR 
Candidates\label{NGC7793RadioSNRsSection}}
Our radio observations of NGC 7793 have yielded five candidate radio
SNRs, to be added to the 28 SNRs previously known through the optical
survey conducted by \citeauthor{BL97}. We present images that depict the
multi-wavelength properties of these candidate radio SNRs in Figures
\ref{fig4} through \ref{fig8}, using a similar format to that used
for the images of the optically-identified SNRs as shown in Figures 
\ref{fig2} and \ref{fig3}. Similar to our search for candidate  
radio SNRs in NGC 300 (\citeauthor{Pannuti00}), these sources have been 
chosen based on their non-thermal radio emission and their 
association with regions of H$\alpha$ emission in NGC 7793.
Extensive optical confusion in these regions may have prevented the
prior detection of these sources by optical methods. We find that the
radio emission from some of these candidate radio SNRs is confused with 
diffuse thermal
emission from the region of H$\alpha$ emission itself. This affects two
of our five candidate radio SNRs -- namely NGC 7793-R3 and NGC 7793-R4 --
and is similar to the situation we encountered in our previous study of
candidate radio SNRs in NGC 300. Because of the larger distance to NGC 7793
compared to NGC 300, our survey for candidate radio SNRs has a corresponding
reduction in sensitivity, and only the more luminous candidate radio SNRs 
in this galaxy are detected by our survey. 
\par
Following the paradigm set in our analysis of candidate radio SNRs in
NGC 300 (\citeauthor{Pannuti00}), we have sought coincidences between our
sample of radio sources and the list of X-ray sources identified in this
galaxy by \citeauthor{RP99}. We have found one such coincidence: the candidate
radio SNR NGC 7793-R3 is coincident with the X-ray source P10 as 
identified by \citeauthor{RP99}, with an offset of only $\sim$ 5$\arcsec$. 
This offset is well within the typically-quoted spatial resolution of  
the $\it{PSPC}$ instrument, namely a 90\% enclosed energy radius at
1 keV of 27$\arcsec$ \citep{H92}. Both
the X-ray source and the radio source correspond to the HII region Hodge
23, as cataloged by \citet{H69}. \citeauthor{RP99} used two thermal 
bremsstrahlung models to fit the spectra of this source, and both of the
resultant fits correspond to spectral energy distributions that peak 
below 1 keV (0.77$\pm$0.64 and 0.63$\pm$0.52 keV, respectively), consistent
with the X-ray spectra expected for an SNR. The luminosity estimates for
this source are approximately 10$^{38}$ ergs sec$^{-1}$ at the assumed
distance to NGC 7793, which is rather large for a single individual
SNR. It is very likely that this fit incorporates diffuse emission in
the neighborhood of the SNR that was not resolved by the $\it{ROSAT}$
$\it{PSPC}$ observations, or it may be that additional X-ray flux is
supplied by other SNRs in the region. We note that a second candidate
radio SNR was found in an adjacent region of H$\alpha$ emission (NGC 7793-R4
in the HII region Hodge 21), but the offset of this second source is
considerable ($\sim$ 15$\arcsec$). Higher resolution X-ray observations
of this source may lead to a clearer interpretation of its properties. 
\par
To complete this discussion on radio-selected SNRs, we consider
background radio sources seen through the disk of NGC 7793 and 
the potential confusing effects such sources may have on this survey.
We repeat two tests to study the possible effects of background
source confusion that we first presented in our previous study of
NGC 300 (\citeauthor{Pannuti00}). First, we use the estimates of \citet{MC85}
for the number density of background radio sources at 20~cm and  
determine that approximately four background sources at that wavelength 
should be detected in random directions through
the disk of NGC 7793 at a 3 $\sigma$ level of 0.2 mJy
or greater. Similarly, a search for position coincidences between
``negative" radio sources that are detected at the -3 $\sigma$ level
with regions of H$\alpha$ emission yielded only one such coincidence,
which is consistent with the result expected from random superpositions
of these background sources with regions of H$\alpha$ emission seen 
in NGC 7793. As we noted with NGC 300, while the criteria for our radio
survey for SNRs selects against isolated radio-emitting SNRs, we do
detect SNRs located within dense regions of H$\alpha$ emission, and 
therefore we can estimate the number of SNRs missed by optical surveys.
\section{Properties of the X-ray-Selected SNR 
Candidates\label{NGC7793XraySNRsSection}}
In order to complete our multi-wavelength search for SNRs in NGC 7793,
we will now discuss the X-ray sources previously described in Section
\ref{NGC7793XrayObsSubSection}. The X-ray source P8 and its association with 
the optically-identified SNR N7793-S26 has already been discussed in
Section \ref{NGC7793OptSNRsSection}. Likewise, the association between the
X-ray source P10 and the candidate radio SNR NGC 7793-R3 has already been
presented in Section \ref{NGC7793RadioSNRsSection}. We will now consider
the five remaining X-ray sources found within the disk of NGC 7793 by
\citeauthor{RP99} to determine if any of these remaining sources may be
candidate X-ray SNRs. We note that it is possible that an X-ray source
seen through the disk of NGC 7793 may actually be a background source
unrelated to the foreground galaxy. To address this possible source of
confusion, we will only consider those X-ray sources which also possess
an optical counterpart within the galaxy to be candidate X-ray SNRs. 
\par
As was the case in our study of X-ray sources in NGC 300 
(\citeauthor{Pannuti00}), we will consider an X-ray source to be a candidate  
X-ray SNR if it satisfies two criteria: first, the X-ray source must 
not show convincing evidence for time variability
in its X-ray luminosity, and second, it must possess a spectrum which 
can be optimally fit using a thermal bremsstrahlung model with a 
characteristic temperature $\it{kT}$ of 1 keV or less. To apply the
first criterion, we note that while analyzing their X-ray data for this
galaxy, \citeauthor{RP99} measured the count rates over the two observing 
epochs 
for each of the 27 sources detected within the field of the $\it{ROSAT}$
$\it{PSPC}$ observations of NGC 7793. 
Based on these count rates, \citeauthor{RP99} also determined the probability
of variability for each source. We reproduce the results of their analysis
in Table \ref{NGC7793XraySources}, and note that for the two
X-ray sources P9 and P13, \citeauthor{RP99} determined a 100\% probability of
variability. On the basis of that finding, we discard those two 
sources from consideration as candidate X-ray SNRs. The two X-ray sources
which we have already associated with SNRs (P8 and P10) both show only
slight evidence for X-ray variability (probabilites of 13\% and 7\%,
respectively).  
As for the second
criterion, we note that the X-ray source P11 was best-fit by \citeauthor{RP99} 
using a power-law model rather than a thermal bremsstrahlung model, and
for that reason P11 was not considered a candidate X-ray SNR.  
\par
The above analysis leaves two X-ray sources to consider, P6 and P7. 
Unfortunately,
it is not a straightforward matter to determine the true nature of 
either of these sources. In our previous analysis of X-ray sources in
NGC 300 (\citeauthor{Pannuti00}),  we could isolate the candidate 
X-ray SNRs on the
basis of spectra which, according to the fits produced by \citet{RPS97},
appeared to favor one spectral-fitting model (power law or thermal
bremsstrahlung) over the other. Inspection of the computed $\chi^2$'s for
each model (calculated to evaluate the quality of each fit) shows that
neither of the two models is distinctly superior to the other for
either of the two sources. Such ambiguity originates from the low 
count rates registered by each source (1.8$\pm$0.4 $\times$ 10$^{-3}$ and 
2.4$\pm$0.5 $\times$ 10$^{-3}$ counts sec$^{-1}$ for P6 and P7, 
respectively), and it prevents the conclusive classification 
of each source. \citeauthor{RP99} argue that these sources
may be background objects unrelated to NGC 7793, on the basis of the      
similarity of their spectra with quasars, the lack of optical 
counterparts in NGC 7793 for both of these sources, and their possible
associations with faint stellar-like objects, with apparent magnitudes
$m$$_V$ of 16 and 18, respectively. The somewhat high likelihood of
variability for these sources (41\% and 37\%, respectively) are also
arguments against classifying these sources as candidate X-ray SNRs.
Inspection of our own radio images fails to reveal sources at either
the 6 or 20~cm wavelengths at the 3$\sigma$ level or higher that are
associated with P6 or P7. A true understanding of these sources must
therefore await additional observations, and we therefore do not 
consider either of these sources to be new candidate X-ray SNRs.
To summarize the results of this section, we have not found any new
candidate X-ray SNRs besides the previously discussed X-ray sources P8 
and P10. 
\section{Discussion\label{DiscussionSection}}
As a result of our X-ray and radio searches for SNRs in NGC 7793,
five new SNRs have been detected, to be added to the list of 28 
optically-identified SNRs found in this galaxy by \citeauthor{BL97}. 
All five of these SNRs are candidate radio SNRs, and one of them --
NGC 7793-R3 -- possesses a counterpart in the X-ray. Another candidate
radio SNR, NGC 7793-R4, may also be associated with this X-ray source.
In Table 
\ref{N7793CrossRefs}, we present a listing of cross-references for
these new SNRs. The surprising 
lack of overlap in detected SNRs found by the three different search
techniques -- X-ray, optical and radio -- is consistent with our
results for NGC 300 (\citeauthor{Pannuti00}). A similar result has been
found by \citet{L01} in the case of the nearby galaxy NGC 6946: those
authors found very limited overlap between the sample of candidate
radio SNRs presented by \citet{LDG97} and the sample of optically-identified
SNRs prsented by \citet{MF97} in that galaxy. In Figure \ref{fig9}, we 
present Venn diagrams to depict the amount of intersection seen
between selected sets of SNRs in the galaxies NGC 300, NGC 7793 and
M33: we have prepared these diagrams for NGC 300 and NGC 7793 based
on the results of \citeauthor{Pannuti00} and the present work, respectively, 
while the diagram for M33 has been created based on results from
\citet{HP01}, \citet{G98} and \citet{GDKGV99}. We note that we have
updated the Venn diagrams for both M33 and NGC 300 from 
\citeauthor{Pannuti00} in light of recent
work on the X-ray source populations in those two galaxies as presented
by \citet{HP01} and \cite{RP01}, respectively. Here we briefly describe 
the impact of these two recent works on the Venn diagrams: a detailed
study of the SNR population in NGC 6946, based on X-ray, optical and
radio observations, as well as a statistical study of the properties
of candidate radio SNRs in nearby galaxies, will be the subjects of
future papers.
\par
In their study of the X-ray population of M33, \citet{HP01} 
prepared a catalog of X-ray sources in that galaxy based on data from
$\it{ROSAT}$ $\it{PSPC}$ and $\it{HRI}$ (High Resolution Imager) 
observations. A total of 184 sources were detected within 50 arcminutes
of the nucleus, an increase in the number of known X-ray sources
in this galaxy as detected by \citet{L96} by over a factor of three.
\citet{HP01} classified sixteen of the 184 sources as SNRs based on
their spectral properties, of which eleven are also detected in the
optical and radio, two are also detected in the optical but not the
radio, and the remaning three are not found at either of the other
two wavelengths. \citet{HP01} noted that only thirteen of the 98
SNRs found by \citet{G98} using optical methods are also detected
in the X-ray, and suggested that the optically-identified SNRs that
lack X-ray emission are likely expanding into a low density ISM. 
\par
Similar to the work presented by \citet{HP01}, \citet{RP01} examined 
$\it{ROSAT}$ $\it{PSPC}$ and $\it{HRI}$ observations of NGC 300 and,
through their analysis, increased the number of $\it{PSPC}$ detected 
sources within the disk of NGC 300 from the fifteen presented by 
\citet{RPS97} to 26. \citet{RP01} refuted the classification  
by \citeauthor{Pannuti00} of two X-ray sources as candidate X-ray
SNRs, namely the sources denoted as P44 and P56 by \citet{RP01}.
P44 was re-classified by those authors as an X-ray binary (thereby calling 
into
doubt its association with the candidate radio SNR R2 as claimed by
\citeauthor{Pannuti00}) while P56 was classified as a foreground star.
\citet{RP01} identified six candidate X-ray SNRs, three of which (P29,
P38 and P49, in their notation) are associated with three optically-identified 
SNRs cataloged by \citeauthor{BL97} (N300-S6, N300-S10 and NGC 300-26,   
respectively). Using these new results and retaining the association
(as argued in \citeauthor{Pannuti00})
between the X-ray source denoted as P41 by \citet{RP01} and the candidate
radio SNR R5, we have revised
the Venn diagram for NGC 300 accordingly. 
\par
Considering these three Venn diagrams and returning to the argument
we presented in \citeauthor{Pannuti00}, we  
feel that the driving effect for such limited intersection is the
distance to the galaxy. The sensitivities at all three wavelengths for 
the surveys in all three galaxies 
are presented in Table \ref{SNRSurveySensitivities}: we note
that we have updated the sensitivities of the X-ray surveys in M33 and
NGC 300 from \citeauthor{Pannuti00}
in light of the recent work presented by \citet{HP01} and 
\citet{RP01}, respectively. Because the sensitivities of the surveys  
for each galaxy are very similar for each wavelength, we argue that
distance plays the most dramatic role in determining the likelihood
of detecting an SNR.  
\par
Furthermore, our results for NGC 7793 represent the third example of
a galaxy (following NGC 300 and NGC 6946) where very few of the SNRs
found by optical methods are also seen in the X-ray and radio. We
interpret this result to suggest that optical surveys favor the
detection of SNRs located in regions of lower density, and may therefore
select the SNRs produced by Type Ia SNe. In contrast, radio and X-ray
surveys for SNRs in a galaxy are biased toward the detection of
SNRs located in regions of high density, and therefore may preferentially
select the SNRs that were most likely produced by Type II SNe. 
Therefore, the density of the ambient ISM surrounding
an SNR is strongly linked to the spectral characteristics of the SNR. 
Additional observations and analysis are needed to more thoroughly
explore these results. 
\section{Conclusions\label{ConclusionsSection}}
We have presented a multi-wavelength search and analysis of the
SNR population in NGC 7793. The results and conclusions of this
work can be summarized as follows: 
\par
1) NGC 7793, a nearby spiral galaxy, has been observed in the X-ray, optical
and radio wavelengths to analyze its resident SNR population. This analysis
has examined X-ray observations of the galaxy made by the ${\it ROSAT}$
${\it PSPC}$ instrument, an H$\alpha$ image of the galaxy and new 6 and
20~cm observations made with the VLA. We have analyzed both the X-ray and
radio spectral properties of the SNRs previously identified in the optical by
\citeauthor{BL97}, and in addition we have searched for new
candidate X-ray and radio SNRs.
\par
2) N7793-S26 is the only optically-identified SNR that possesses both
X-ray and non-thermal radio emission. The extreme radio 
luminosity of this object and its large filamentary structure have led 
to speculation about its creation. A study of the energetics of
this SNR suggest that it was created by multiple supernova explosions
rather than a single supernova event.  
One other optically-identified SNR, N7793-S11, possesses non-thermal 
radio emission but no X-ray emission. Consistent with prior studies
of the SNR populations in the galaxies NGC 300 and NGC 6946,  
X-ray and radio emission from
optically-identified SNRs are in general not detected. This lends more
support to the hypothesis that searches conducted for SNRs at each 
wavelength -- X-ray, optical and radio -- all possess inherent biases.  
We also find additional evidence that the density of the
ambient ISM surrounding an SNR plays a critical
role in dictating the SNR's spectral characteristics. 
\par
3) A search for non-thermal radio sources at 6~cm and 20~cm (with a
minimum detection level of 3$\sigma$) that are close to or within HII  
regions has yielded five candidate radio SNRs. Of these five  
sources, one (NGC 7793-R3) possesses an X-ray counterpart at the
3$\sigma$ level, namely P10, from the listing of X-ray sources in this
galaxy that was prepared by \citeauthor{RP99}. Another candidate radio
SNR, NGC 7793-R4, may also contribute X-ray flux to P10. 
\par
4) A search for candidate X-ray SNRs has not revealed any new sources
in addition to the SNRs already found through radio and optical surveys.
The search for SNRs at this wavelength is complicated by the presence
of considerable diffuse X-ray emission throughout the entire disk of 
this galaxy. 
\par
5) The multi-wavelength campaign has added five new candidate SNRs to
the 28 previously identified through optical methods. While the number
of new detected SNRs is noticeably lower than the number of new SNRs
found in our study of NGC 300 as presented in \citeauthor{Pannuti00},   
we feel that this result is linked to the larger distance to NGC 7793
than NGC 300. Because the X-ray and radio observations of NGC 7793 
did not have improved sensitivities to compensate for the increased distance
to this galaxy compared to NGC 300, both surveys can only sample more
luminous portions of the X-ray and radio SNR population in NGC 7793
compared to NGC 300.
\begin{acknowledgements}
This paper is derived in large part from the Ph.D dissertation of 
T. G. P., conducted at the University of New Mexico. We thank
the referee, Schuyler Van Dyk, for many useful comments and 
suggestions which have greatly improved this work. T. G. P. acknowledges 
fruitful discussions with Glenn E. Allen and Miroslav Filipovi\'{c}. 
This research has made use of NASA's Astrophysics Data System Abstract
Service and the NASA/IPAC Extragalactic Database (NED) 
which is operated by the Jet Propulsion Laboratory, California Institute
of Technology, under contract with the National Aeronautics and Space
Administration. The {\it ROSAT} project is supported by the German
Bundesministerium f\"{u}r Bildung und Forschung (BMBF) and the 
Max-Planck-Gesellschaft (MPG). T. G. P. gratefully acknowledges
support from both the {\it Chandra} X-ray Center High Energy
Transmission Grating (HETG) group through grant NAS8-38249 and the
NRAO.  
\end{acknowledgements}

\clearpage
%
%
\begin{figure}
\figurenum{4}
\epsscale{0.95}
\plottwo{f4a.eps}{f4b.eps}
\plotone{f4c.eps}
\caption{Same as Fig. 2, but for the candidate radio SNR  
NGC 7793-R1 in the HII region Hodge 31. In Fig. 4b, the contours are placed 
at -3, -2, 3, 4, 5 and 6 times the rms noise level at 6~cm, and in Fig.  
4c, the contours are placed at -3, -2, 3 and 3.5 times the rms noise level
at 20~cm. See $\S$ \ref{NGC7793RadioSNRsSection}. \label{fig4}} 
\end{figure}
\begin{figure}
\figurenum{5}
\epsscale{0.95}
\plottwo{f5a.eps}{f5b.eps}
\plotone{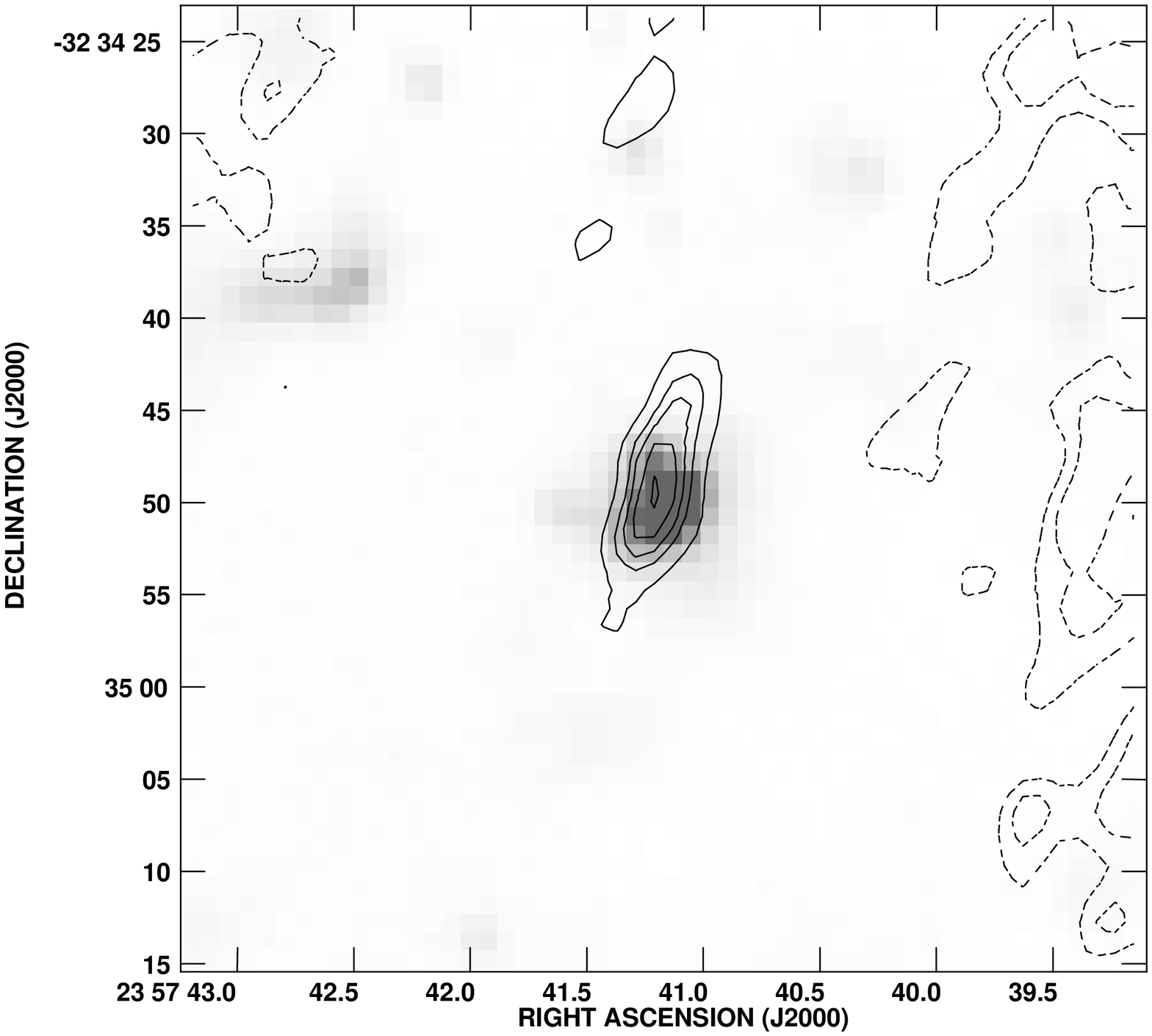}
\caption{Same as Fig. 2, but for the candidate radio SNR  
NGC 7793-R2 in the HII region Hodge 27. In Fig. 5b, the contours are placed 
at -3, -2, 3, 4, 5, 6 and 7 times the rms noise level at 6~cm, and in Fig.  
5c, the contours are placed at -3, -2, 3, 4, 4.5, 5 and 5.5 times the rms 
noise level at 20~cm. See $\S$ \ref{NGC7793RadioSNRsSection}. \label{fig5}} 
\end{figure}
\begin{figure}
\figurenum{6}
\epsscale{0.95}
\plottwo{f6a.eps}{f6b.eps}
\plotone{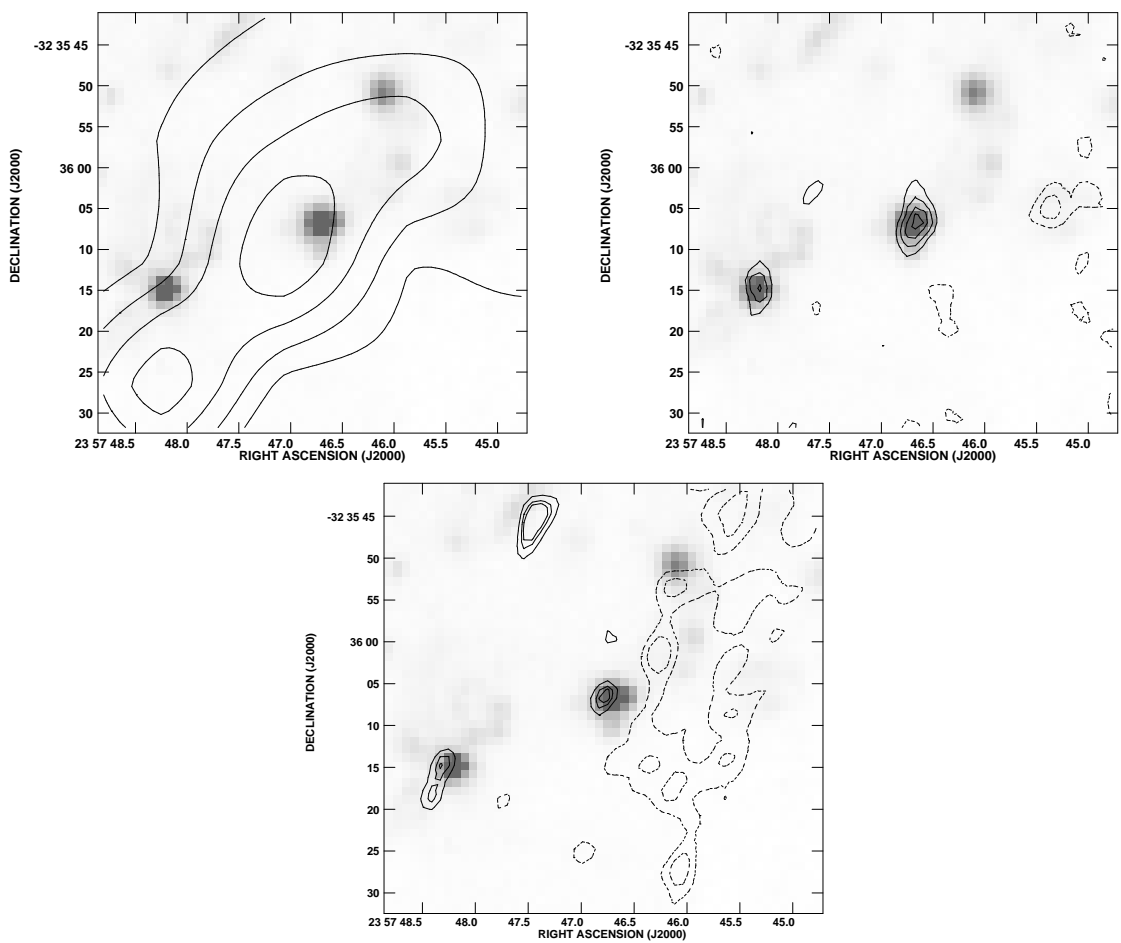}
\caption{Same as Fig. 2, but for the candidate radio SNR  
NGC 7793-R3 in the HII region Hodge 23. In Fig. 6a, notice the
proximity to the X-ray source P10. In Fig. 6b, the contours are placed
at -3, -2, 3, 4, 5 and 5.75 times the rms noise level at 6~cm, and in Fig.
6c, the contours are placed at -3, -2, 3, 3.5 and 3.75 times the rms
noise level at 20~cm. See $\S$ \ref{NGC7793RadioSNRsSection}. \label{fig6}} 
\end{figure}
\begin{figure}
\figurenum{7}
\epsscale{0.95}
\plottwo{f7a.eps}{f7b.eps}
\plotone{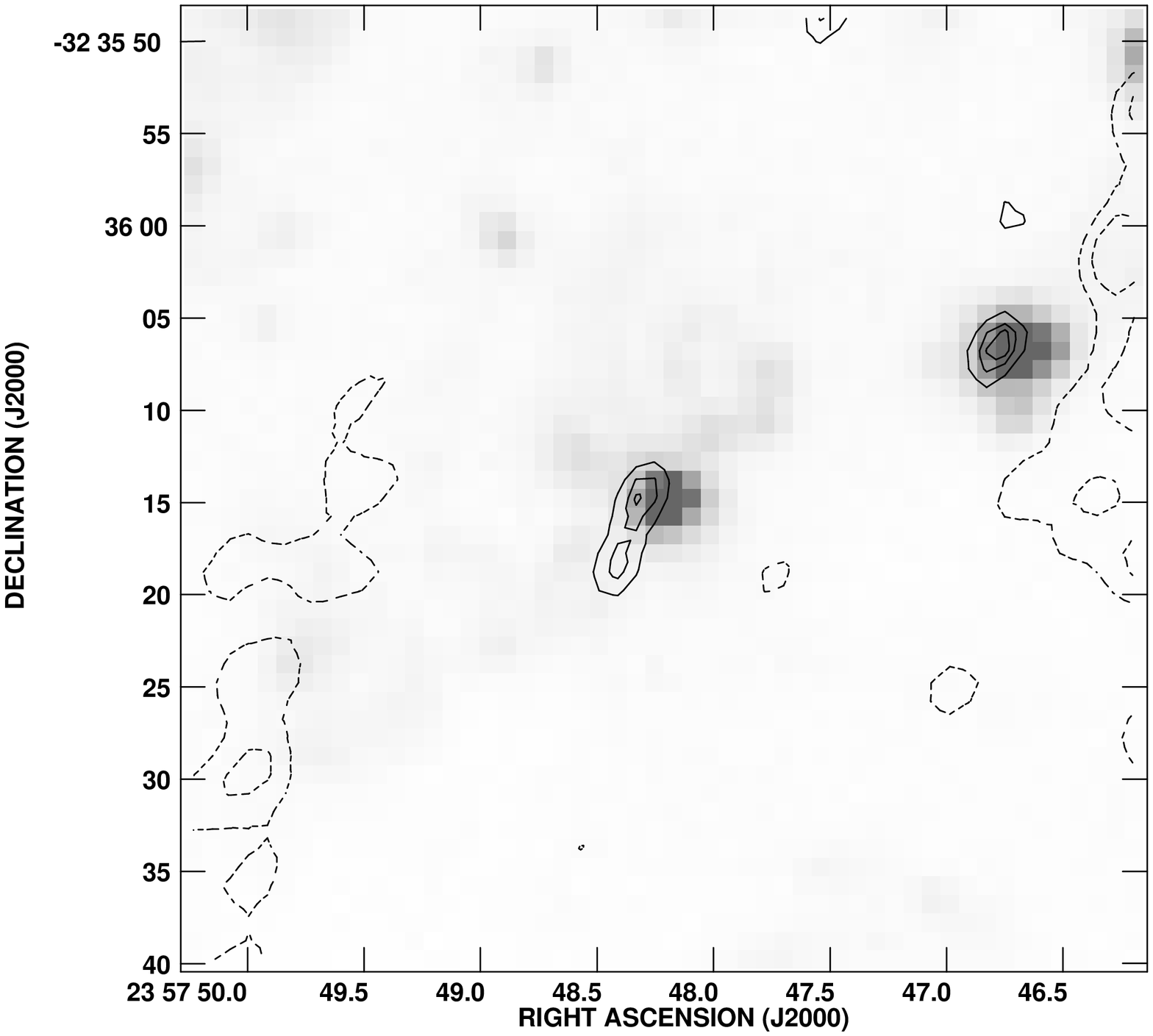}
\caption{Same as Fig. 2, but for the candidate radio SNR  
NGC 7793-R4 in the HII region Hodge 21. In Fig. 7b, the contours are placed
at -3, -2, 3, 4, 5 and 5.75 times the rms noise level at 6~cm, and in Fig.
7c, the contours are placed at -3, -2, 3, 3.5 and 3.75 times the rms
noise level at 20~cm. See $\S$ 
\ref{NGC7793RadioSNRsSection}. \label{fig7}}
\end{figure}
\begin{figure}
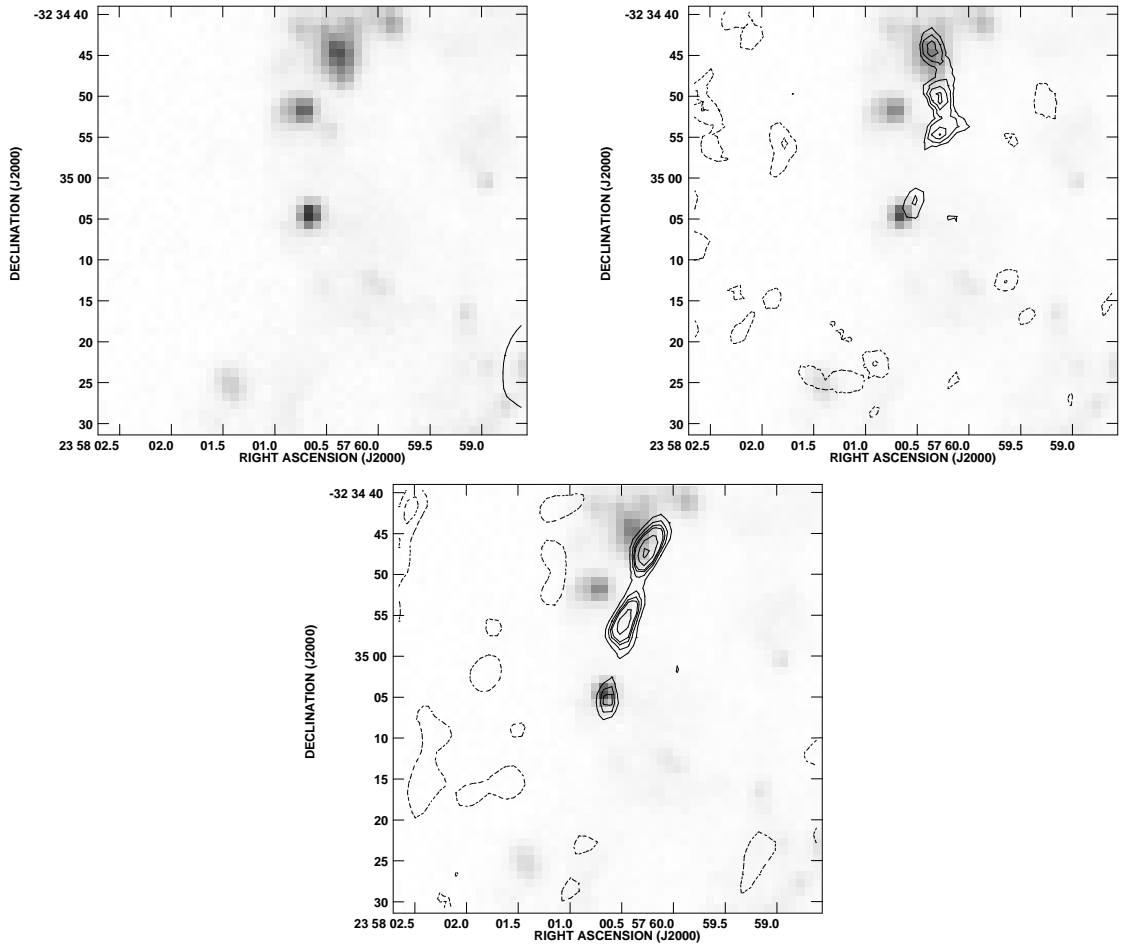

\figurenum{8}
\epsscale{0.95}
\plottwo{f8a.eps}{f8b.eps}
\plotone{f8c.eps}
\caption{Same as Fig. 2, but for the candidate radio SNR  
NGC 7793-R5 in an uncataloged HII region. In Fig. 8b, the contours are placed 
at -3, -2, 3, 3.5, 4 and 4.5 times the rms noise level at 6~cm, and in Fig.
8c, the contours are placed at -3, -2, 3, 3.8, 4, 4.5 and 5 times the rms
noise level at 20~cm. See $\S$ 
\ref{NGC7793RadioSNRsSection}. \label{fig8}}
\end{figure}
\begin{figure}
\figurenum{9}
\epsscale{0.95}
\plottwo{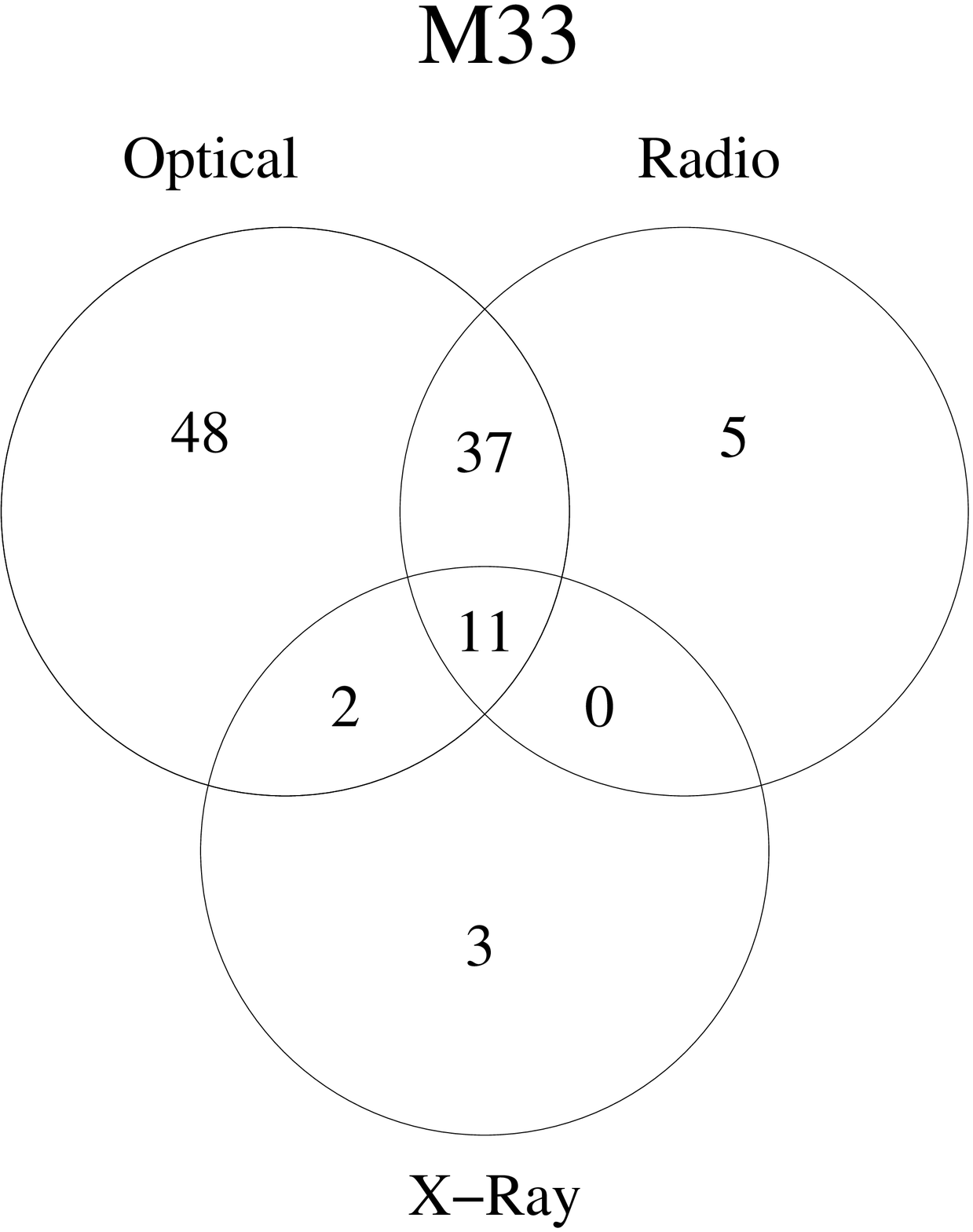}{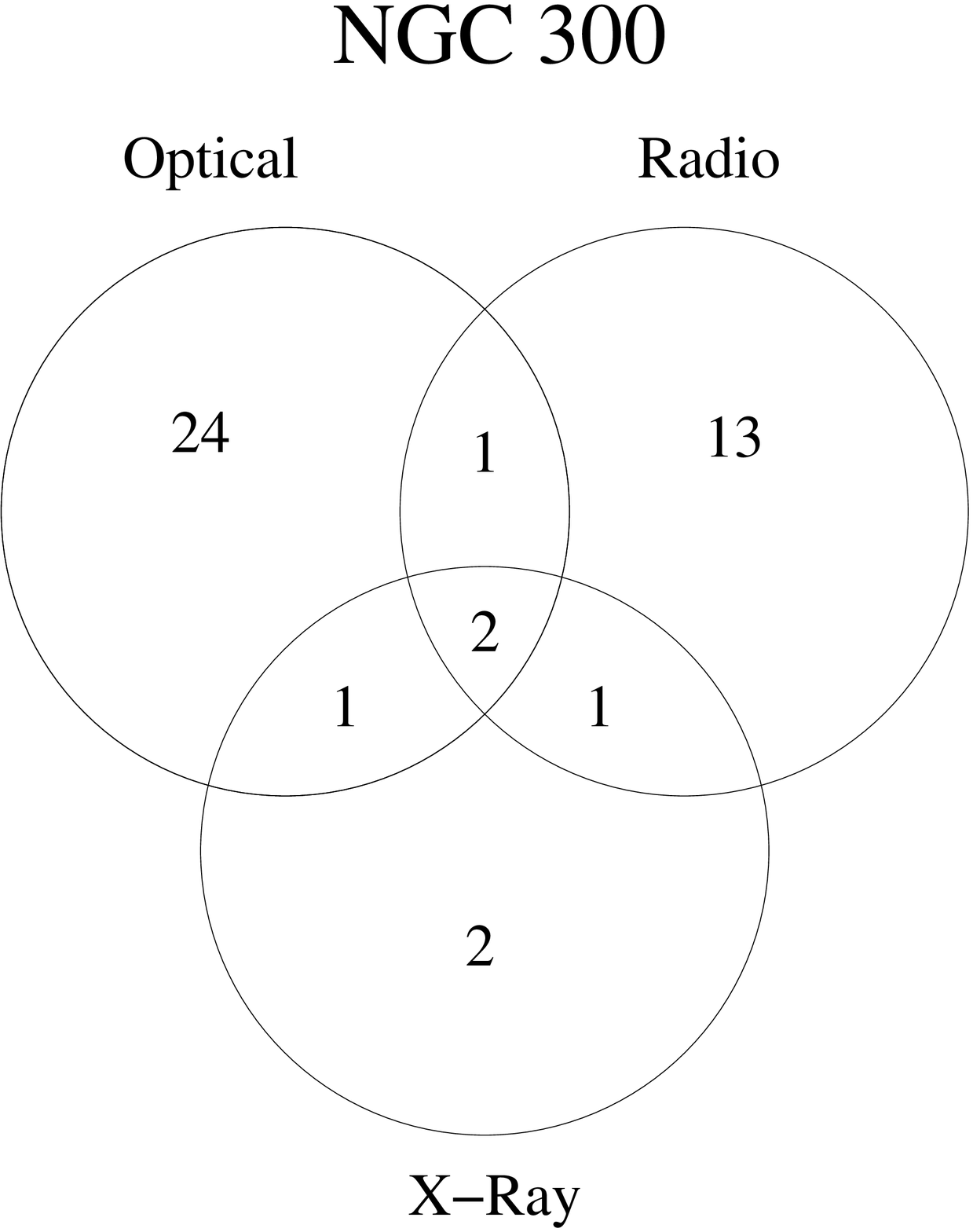}
\plotone{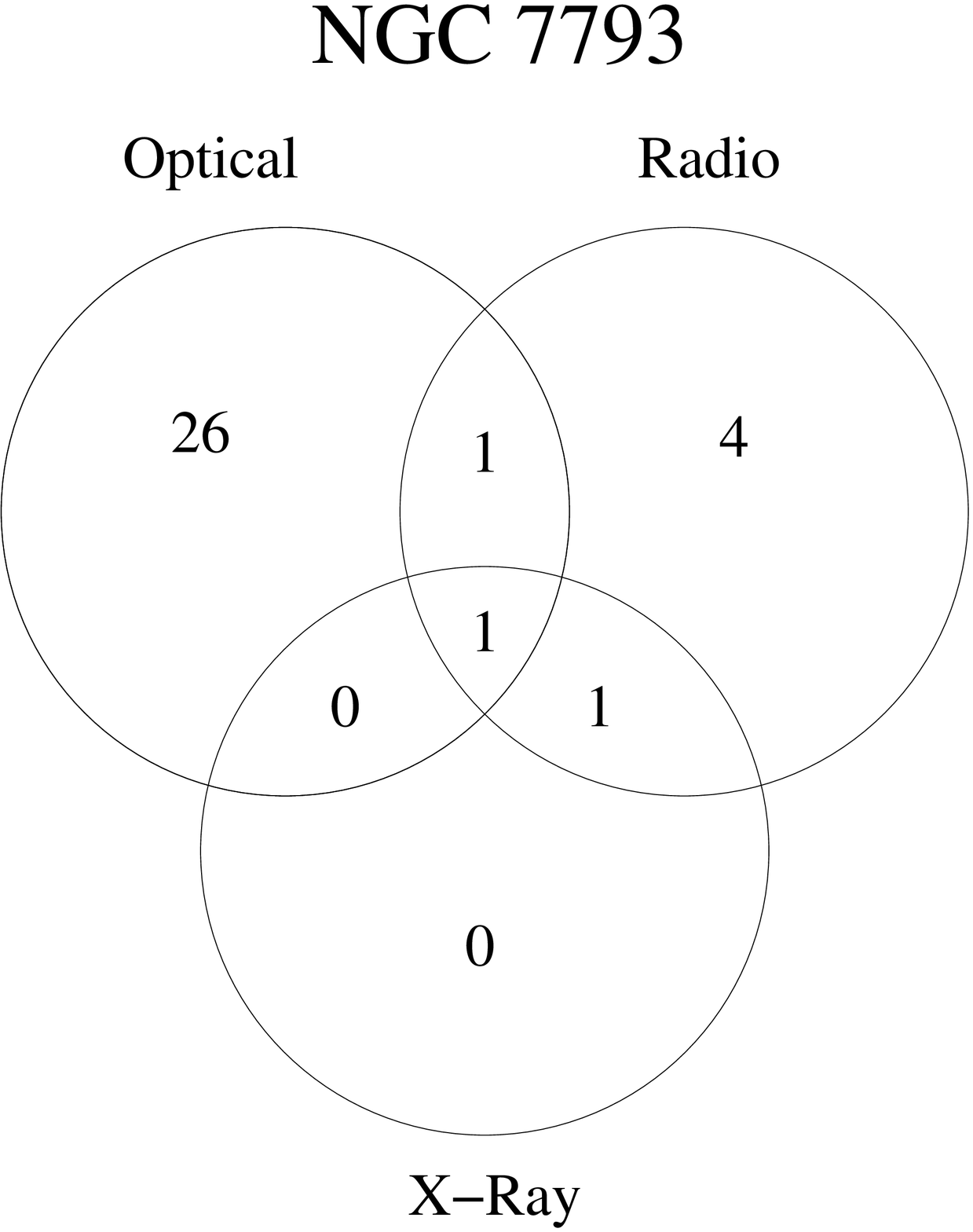}
\caption{Venn Diagrams depicting the intersection of 
selected sets of SNRs (X-ray, optical and radio) for the galaxies M33,
NGC 300 and NGC 7793. The galaxies have been arranged in order of
increasing distance. Notice how limited the intersection between the
three sets becomes with increasing distance to the galaxy. \label{fig9}}  
\end{figure}
%
%
\begin{deluxetable}{lc}
\tablecaption{General Properties of NGC 7793\label{7793Props}}
\tablewidth{0pt}
\tablehead{
\colhead{Property} & \colhead{NGC 7793}  
}
\startdata
R.A. (J2000.0) & 23h 57m 49.2s\tablenotemark{a}\\
Dec  (J2000.0) & $-$32$^{\circ}$ 35$\arcmin$ 
24$\arcsec$\tablenotemark{a}\\
Galactic Latitude & $-$77.17\tablenotemark{a}\\
Galactic Longitude & 4.52\tablenotemark{a}\\
Observed Diameter $\it{D_{25}}$ (arcmin)  & 9.2\tablenotemark{b}\\
Axial Ratio $\it{d/D}$ & 0.70\tablenotemark{b}\\
N$_H$ Column Density (cm$^{-2}$) & 1.14$\times$10$^{20}$\tablenotemark{c}\\
Hubble Type & SA(s)d\tablenotemark{a}\\
Distance (Mpc) & 3.38\tablenotemark{d} \\
Inclination $\it{i}$ ($^{\circ}$)& 50\tablenotemark{b}\\
\enddata  
\tablenotetext{a}{NED Database.}
\tablenotetext{b}{\citet{T88}.}
\tablenotetext{c}{\citeauthor{RP99}.}
\tablenotetext{d}{\citet{PC88}.}
\end{deluxetable}
\clearpage
%
%
\begin{deluxetable}{lcccccc}
\tablecaption{Parameters of the NGC 7793 Observations
with the VLA\label{VLAObs}}
\tablewidth{0pt}
\tablehead{
&& \colhead{Field} && \colhead{Linear} & \colhead{Synthesized} & 
\colhead{Position}\\ 
& \colhead{Frequency} & \colhead{of View} & \colhead{Sensitivity} & 
\colhead{Resolution} & \colhead{Beam} & \colhead{Angle}\\ 
\colhead{Configuration} & \colhead{(GHz)} & \colhead{(arcmin)} & 
\colhead{($\mu$Jy)} & \colhead{(pc)} & \colhead{(arcsec$^2$)} & 
\colhead{(degrees)}  
}
\startdata
CnB & 4.86 & 9 & 32 & 97 & 7.8 $\times$ 4.2 & $-$3.63\\
BnA & 1.47 & 27 & 60 & 97 & 9.4 $\times$ 4.1 & $-$10.25\\
\enddata
\end{deluxetable}
\clearpage
%
%
\begin{deluxetable}{ccccccccc}
\tablecaption{Radio-Selected SNR Candidates in NGC 7793\label{N7793RadioSNRs}}
\tablewidth{0pt}
\tablehead{
\colhead{Source} & \colhead{RA} & \colhead{Dec} & \colhead{\ssix} & 
\colhead{\dssix} & \colhead{\stwenty} & \colhead{\dstwenty}\\  
\colhead{Name} & \colhead{(J2000.0)} & \colhead{(J2000.0)} & 
\colhead{(mJy)} & \colhead{(mJy)} & \colhead{(mJy)} & \colhead{(mJy)} & 
\colhead{$\alpha$\tablenotemark{a}} & \colhead{\da} 
}
\startdata
N7793-S11 & 23 57 47.3 & $-$32 35 23 & 0.21 & 0.10 & 0.45 & 0.15 & 
$+$0.6 & 0.5\\
N7793-S26\tablenotemark{b} & 23 58 00.0 & $-$32 33 19 & 1.24 & 0.19 & 3.75 &
0.30 & $+$0.9 & 0.2\\
NGC 7793-R1 & 23 57 40.2 & $-$32 37 38 & $<$\tablenotemark{c} & 0.10 
& 0.22 & 0.07 & $>$0.7 & -- \\ 
NGC 7793-R2 & 23 57 41.2 & $-$32 34 50 & 0.24 & 0.08 & 0.39 & 0.13 & $+$0.4
& 0.4\\
NGC 7793-R3 & 23 57 46.8 & $-$32 36 07 & 0.22\tablenotemark{d}    
& 0.04 & 0.26 & 0.07 & $+$0.2 & 0.3\\
NGC 7793-R4 & 23 57 48.4 & $-$32 36 15 & 0.16\tablenotemark{d} & 0.04 & 
0.22 
& 0.07 & $+$0.3 & 0.3\\
NGC 7793-R5 & 23 58 00.7 & $-$32 35 06 & 0.12 & 0.04 & 0.25 & 0.07 & 
$+$0.6
& 0.4\\
\enddata
\tablecomments{Units of right ascension are hours, minutes and seconds,
and units of declination are degrees, arcminutes and arcseconds.}
\tablenotetext{a}{S$_{\nu}$ $\propto$ $\nu$$^{-\alpha}$.}
\tablenotetext{b}{Here we refer to the combination of the two sources
denoted as ``N7793-S26" and ``N7793-S26ext" as one source.} 
\tablenotetext{c}{Not detected at the 3$\sigma$ level at this wavelength.}
\tablenotetext{d}{Diffuse radio emission from the adjacent HII region makes 
isolating the radio emission associated with the SNR difficult at this 
wavelength.}
\end{deluxetable}
\clearpage
%
%
\begin{deluxetable}{cccccccc}
\tabletypesize{\scriptsize}
\tablecaption{Properties of X-ray Sources in 
NGC 7793\label{NGC7793XraySources}} 
\tablewidth{0pt}
\tablehead{
&&&& \colhead{Flux} &&&\\
\colhead{Source} & \colhead{RA} & \colhead{Dec} & \colhead{Net} & 
\colhead{(10$^{-14}$ ergs} & \colhead{Hardness} & \colhead{Probability} &  
\colhead{Optical/Radio}\\ 
\colhead{Number} & \colhead{(J2000.0)} & \colhead{(J2000.0)} & 
\colhead{Counts} & \colhead{cm$^{-2}$ sec$^{-1}$)} & \colhead{Ratio}
& \colhead{of Variability} & \colhead{Counterparts}
}
\startdata
P6  & 23 57 48.5 & $-$32 32 30 & 41.3$\pm$9.6   & 2.3$\pm$0.5
& $+$0.1$\pm$0.2 & 41\% & -- \\
P7  & 23 57 52.6 & $-$32 33 11 & 54.7$\pm$10.1  & 3.0$\pm$0.6
& $+$0.2$\pm$0.2 & 37\% & -- \\
P8  & 23 57 59.8 & $-$32 33 24 & 69.5$\pm$12.0  & 3.9$\pm$0.7
& $+$0.1$\pm$0.2 & 13\% & N7793-S26\\
P9  & 23 58 08.6 & $-$32 34 03 & 110.1$\pm$12.5 & 6.2$\pm$0.7
& $>$0.5 & 100\% & -- \\
P10 & 23 57 47.1 & $-$32 36 05 & 173.4$\pm$18.1 & 9.7$\pm$1.0
& $+$0.1$\pm$0.1 & 7\% & H23, R3\\
& & & & & & & (H21?, R4?)\\ 
P11 & 23 58 03.1 & $-$32 36 34 & 41.2$\pm$9.3   & 2.3$\pm$0.5
& $>$0.2 & 37\% & -- \\
P13 & 23 57 51.2 & $-$32 37 23 & 976.0$\pm$32.0 & 54.9$\pm$1.8
& $+$0.9$\pm$0.0 & 100\% & -- \\
\enddata
\end{deluxetable}
\clearpage
\begin{deluxetable}{cccccc} 
\tablecaption{Corrected Positions of Optically-Identified SNRs in NGC
7793\label{N7793OptSNRsPositions}}
\tablewidth{0pt}
\tablehead{
\colhead{SNR} & \colhead{RA} & \colhead{Dec} & \colhead{SNR} & 
\colhead{RA} & \colhead{Dec} \\ 
& \colhead{(J2000.0)} & \colhead{(J2000.0)} && \colhead{(J2000.0)} &
\colhead{(J2000.0)} 
}
\startdata
N7793-S1  & 23 57 38.59 & $-$32 34 37.4 & N7793-S16 & 23 57 54.57 & $-$32 35 
12.1\\ 
N7793-S2  & 23 57 38.66 & $-$32 33 19.0 & N7793-S17 & 23 57 55.40 & $-$32 33 
52.9\\ 
N7793-S3  & 23 57 39.11 & $-$32 35 38.5 & N7793-S18 & 23 57 55.40 & $-$32 34 
33.4\\
N7793-S4  & 23 57 40.61 & $-$32 35 51.7 & N7793-S19 & 23 57 55.86 & $-$32 37 
21.5\\ 
N7793-S5  & 23 57 41.02 & $-$32 37 02.6 & N7793-S20 & 23 57 56.07 & $-$32 37
17.5\\
N7793-S6  & 23 57 43.78 & $-$32 35 27.1 & N7793-S21 & 23 57 56.62 & $-$32 36 
09.8\\ 
N7793-S7  & 23 57 44.05 & $-$32 36 40.6 & N7793-S22 & 23 57 58.50 & $-$32 35 
23.1\\
N7793-S8  & 23 57 44.96 & $-$32 37 41.3 & N7793-S23 & 23 57 59.03 & $-$32 36 
46.7\\
N7793-S9  & 23 57 44.99 & $-$32 37 35.1 & N7793-S24 & 23 57 59.34 & $-$32 36 
06.7\\
N7793-S10 & 23 57 45.76 & $-$32 35 00.7 & N7793-S25 & 23 57 59.83 & $-$32 36 
19.0\\ 
N7793-S11 & 23 57 47.26 & $-$32 35 23.1 & N7793-S26 & 23 58 00.27 & $-$32 33 
22.1\\  
N7793-S12 & 23 57 48.30 & $-$32 36 56.0 & N7793-S26ext & 23 57 59.79 & $-$32 
33 16.0\\  
N7793-S13 & 23 57 51.26 & $-$32 36 32.2 & N7793-S27 & 23 58 06.75 & $-$32 35 
12.5\\ 
N7793-S14 & 23 57 52.58 & $-$32 33 52.5 & N7793-S28 & 23 58 06.75 & $-$32 35 
36.7\\  
N7793-S15 & 23 57 54.36 & $-$32 34 02.6\\ 
\enddata
\end{deluxetable}
\clearpage
\begin{deluxetable}{lcccc}
\tablecaption{Cross-References of All New Candidate Radio SNRs in NGC
7793\label{N7793CrossRefs}}
\tablewidth{0pt}
\tablehead{
\colhead{Candidate} & \colhead{RA} & \colhead{Dec} & 
\colhead{Optical} & \colhead{X-ray} \\ 
\colhead{Radio SNR\tablenotemark{a}} & \colhead{(J2000.0)} & 
\colhead{(J2000.0)} & \colhead{Counterpart?} & \colhead{Counterpart?} 
}
\startdata
NGC 7793-R1  & 23 57 40.2 & $-$32 37 38 & H31 & --\\
NGC 7793-R2  & 23 57 41.2 & $-$32 34 50 & H27 & --\\
NGC 7793-R3  & 23 57 46.8 & $-$32 36 07 & H23 & P10\\
NGC 7793-R4  & 23 57 48.4 & $-$32 36 15 & H21 & (P10?)\\
NGC 7793-R5  & 23 58 00.4 & $-$32 35 06 & Uncataloged & --\\
& & & HII Region & \\
\enddata
\tablenotetext{a}{In this table, the radio counterparts are the radio sources
presented in Table \ref{N7793RadioSNRs}, the optical counterparts are the 
HII regions cataloged by \citet{H69}, and the X-ray counterparts are the 
X-ray sources listed in Table \ref{NGC7793XraySources}.} 
\end{deluxetable}
\clearpage
\begin{deluxetable}{lcccccc}
\tablecaption{Sensitivities of SNR Surveys of M33, NGC 300 and NGC 
7793\label{SNRSurveySensitivities}} 
\tablewidth{0pt}
\tablehead{
\colhead{Galaxy} & \colhead{X-ray} & \colhead{Reference} & 
\colhead{H$\alpha$} & \colhead{Reference} & \colhead{20~cm} & 
\colhead{Reference} \\
& \colhead{(counts} & & \colhead{(ergs cm$^{-2}$} & & \colhead{($\mu$Jy per} \\ 
& \colhead{sec$^{-1}$ arcmin$^{-2}$)} & & \colhead{sec$^{-1}$ arcsec$^{-2}$)} 
& & \colhead{beam)} 
}
\startdata
M33      & 9 $\times$ 10$^{-4}$ & (1) & $\sim$6 $\times$ 10$^{-17}$ & (2) 
& 50 & (3)\\
NGC 300  & 1.2 $\times$ 10$^{-3}$ & (4) & $\sim$1 $\times$ 10$^{-16}$ & (5) & 
60 & (6)\\  
NGC 7793 & 1.7 $\times$ 10$^{-3}$ & (7) & $\sim$1 $\times$ 10$^{-16}$ & (5) & 
60 & (8)\\ 
\enddata
\tablecomments{The references used in this table are as follows: (1)
\citet{HP01}, (2) \citet{G98}, (3) \citet{GDKGV99}, (4) \citet{RP01}, (5)
\citeauthor{BL97}, (6) \citeauthor{Pannuti00}, (7) \citeauthor{RP99} and 
(8) This work.} 
\end{deluxetable}

\begin{thebibliography}{}
\bibitem[Allen et al.(1997)]{A97} Allen, G.E., et al. 1997, \apj, 487, L97
\bibitem[Bartunov et al.(1994)]{B94} Bartunov, O.S., Tsvetkov, D. Yu. \& 
Filimonova, I.V. 1994, \pasp, 106, 1276
\bibitem[BL97(1997)Blair \& Long]{BL97} Blair, W.P. \& Long, K.S.  1997,
\apjs, 108, 261 (BL97) 
\bibitem[Carignan(1985)]{C85} Carignan, C.  1985, \apjs, 58, 107
\bibitem[Carignan \& Puche(1990)]{CP90} Carignan, C. \& Puche, D.  1990,
\aj, 100, 394
\bibitem[D'Odorico et al.(1980)]{DDB80} D'Odorico, S., Dopita, 
M.A. \& Benvenuti, P.  1980, \aaps, 40, 67  
\bibitem[Fabbiano et al.(1992)]{F92} Fabbiano, G., Kim, D.-W. \& 
Trinchieri, G.  1992, \apjs, 80, 531
\bibitem[Ferguson et al.(1996)]{FWGH96} Ferguson, A.M.N., Wyse, R.F.G., 
Gallagher, J.S., III \& Hunter, D.A.  1996, \aj, 111, 2265 
\bibitem[Freedman et al.(1992)]{FMHHMS92} Freedman, W.L., Madore, B.F., 
Hawley, S.L., Horowitz, I.K., Mould, J. \& Sallmen, S.  1992, \apj, 396, 80 
\bibitem[Gordon et al.(1998)]{G98} Gordon, S.M., Kirshner, R.P., Long, K.S.,
Blair, W.P., Duric, N. \& Smith, R.C.  1998, \apjs, 117, 89
\bibitem[Gordon et al.(1999)]{GDKGV99} Gordon, S.M., Duric, N., Kirshner, 
R.P., Goss, W.M. \& Viallefond, F. 1999, \apjs, 120, 247
\bibitem[Greenawalt et al.(1998)]{GWTH98} Greenawalt, B.E., Walterbos,
R.A.M., Thilker, D. \& Hoopes, C.G.  1998, \apj, 506, 135
\bibitem[Haberl \& Pietsch(2001)]{HP01} Haberl, F. \& Pietsch, W. 2001,
\aap, 373, 438 
\bibitem[Harnett(1986)]{H86} Harnett, J.I.  1986, PASAu, 6, 325
\bibitem[Hasinger et al.(1992)]{H92} Hasinger, G., Turner, T.J.,
George, I.M. \& Boese, G. 1992, OGIP Calibration Memo CAL/ROS/92-001 
\bibitem[Hodge(1969)]{H69} Hodge, P.W.  1969, \apjs, 157, 73
\bibitem[Hoopes et al.(1996)]{HWG96} Hoopes, C.G., Walterbos, R.A.M. 
\& Greenawalt, B.E.  1996, \aj, 112, 1429
\bibitem[Hwang et al.(2000)]{HPH00} Hwang, U., Petre, R. \& Hughes, J. P.
2000, \apj, 532, 970
\bibitem[Lacey et al.(1997)]{LDG97} Lacey, C.K., Duric, N. \& 
Goss, W.M.  1997, \apjs, 108, 417
\bibitem[Lacey \& Duric(2001)]{L01} Lacey, C.K. \& Duric, N. 2001,
\apj, 560, 719   
\bibitem[Lacey et al.(2001)]{LLKDBD01} Lacey, C.K., Lazio, T.J.W., 
Kassim, N.E., Duric, N., Briggs, D.S. \& Dyer, K. K. 2001, \apj, 559, 
954
\bibitem[Long et al.(1996)]{L96} Long, K.S., Charles, P.A., Blair, W.P.
\& Gordon, S.M. 1996, \apj, 466, 750 
\bibitem[Matonick \& Fesen(1997)]{MF97} Matonick, D.M. \& Fesen, R.A.
1997, \apjs, 112, 49
\bibitem[Matonick et al.(1997)]{M97} Matonick, D.M., Fesen, R.A., Blair,
W.P. \& Long, K.S. 1997, \apjs, 113, 333 
\bibitem[Mitchell \& Condon(1985)]{MC85} Mitchell, K.J. \& Condon, J.J.
1985, \aj, 90, 1957
\bibitem[Paper I(2000)Pannuti et al.]{Pannuti00} Pannuti, T.G., Duric, N., 
Lacey, C.K., Goss, W.M., Hoopes, C.G., Walterbos, R.A.M. \& Magnor, M. 
2000, \apj, 544, 780 (Paper I)   
\bibitem[Puche \& Carignan(1988)]{PC88} Puche, D. \& Carignan, C.  1988, 
\aj, 95, 1025
\bibitem[Read et al.(1997)]{RPS97} Read, A.M., Ponman, T.J. 
\& Strickland, D.K.  1997, \mnras, 286, 626
\bibitem[RP99(1999)Read \& Pietsch]{RP99} Read, A.M. \& Pietsch, W.  1999,
\aap, 341, 8 (RP99) 
\bibitem[Read \& Pietsch(2001)]{RP01} Read, A.M. \& Pietsch, W. 2001, 
\aap, 373, 473
\bibitem[Schulz(1999)]{S99} Schulz, N.S. 1999, \apj, 511, 304 
\bibitem[Tr\"{u}mper(1992)]{T92} Tr\"{u}mper, J.  1992, QJRAS, 33, 165
\bibitem[Tully(1988)]{T88} Tully, R.  1988, $\it{Nearby}$ $\it{Galaxies}$
$\it{Catalog}$ (Cambridge: Cambridge University Press)
\end{thebibliography}
\end{document}